\documentclass[12pt,margins=1in]{iopart}

\usepackage{graphicx,subcaption}
\usepackage{tikz}
\usepackage{tcolorbox}

\usepackage{xcolor}
\usepackage[colorlinks,linkcolor=blue, citecolor = blue, urlcolor = blue]{hyperref}
\usepackage{cleveref}
\usepackage{multicol}
\usepackage{tablefootnote} 
\usepackage{multirow}

\usepackage{iopams}  
\usepackage{lmodern}
\usepackage{orcidlink}
\usepackage{chemfig}
\begin{document}

\title[]{\textbf{Electron impact partial ionization cross section and thermal rate coefficients of gaseous refrigerants}}

\author{Suriyaprasanth S  \orcidlink{0000-0002-3939-1446} and Dhanoj Gupta \thanks{corresponding author} \orcidlink{0000-0001-6717-8194}}
\address{Department of Physics, School of Advanced Sciences, Vellore Institute of Technology, Katpadi, Vellore - 632014, Tamil Nadu, India}
\ead{suriyaprasanth.s@vit.ac.in, dhanoj.gupta@vit.ac.in}


\begin{abstract}
We have calculated the electron impact partial and total ionization cross sections of important gaseous targets, such as Trifluoromethane (CHF$_3$), 1,1,1,2-Tetrafluoroethane $(\mathrm{C_2H_2F_4})$ or R134a, 1,1,1-Trifluoroethane $(\mathrm{C_2H_3F_3})$ or R143a, 1,1,1-Trifluoropropane $(\mathrm{C_3H_5F_3})$ or R263fb, and 3,3,3-Trifluoropropene $(\mathrm{C_3H_3F_3})$ or R1243zf using the binary encounter Bethe model and its variants. The corresponding rate coefficients are calculated for total and partial ionization cross sections using the Maxwell's velocity distribution function. Our data for partial and total ionization along with the rate coefficient showed good agreement with the existing data in the literature. The targets studied are important for plasma applications and are used in gas-based detectors at high-energy physics experiments.
\end{abstract}

%
\vspace{2pc}
\submitto{$\rm{arXiv}$}
%
\maketitle
%
\ioptwocol

\section{Introduction}\label{sec1}

The study of electron and positron scattering is of fundamental importance in many areas of physics and engineering such as plasma modelling, aeronomy, studying radiation damage in living cells, modelling electron transport in the matter, and many others \cite{alves2018foundations,fox1996aeronomy,sanche2005low,zein2021electron}. In the present work, we mainly focus on exploring the electron inelastic scattering on neutral molecules used as refrigerants in household appliances and as gaseous detectors in high-energy physics experiments. These refrigerants used in gas-based detectors are inevitable as the gases containing fluorine atoms provide optimal performance while detecting elementary particles \cite{Saviano_2018}. However, researchers are now striving to eliminate the use of these gases due to their severe environmental impact. For example, experiments like the mini-iron calorimeter (mini-ICAL), which uses resistive plate chambers (RPC), rely on a quenching medium mixture of R134a, isobutane, and SF$_6$ to achieve precise time resolution and efficient avalanche operation \cite{john2023simulation}. Similarly, the Compact Muon Solenoid (CMS) experiment at the CERN also uses an RPC for the detection of charged particles which also uses the same R143a mixture in their RPCs. Although R143a gas is suitable for detectors, its global warming potential (GWP) is quite high, raising concern for its usage \cite{Saviano_2018}. 
To certify a gas as a viable alternative, we must examine its electron swarm parameters, effective ionization coefficients, transport properties such as drift velocity, Townsend coefficients, attachment coefficients and both longitudinal and transverse diffusion coefficients that can be studied by numerically solving the Boltzmann's equation or using the Monte-Carlo simulation. Popular and well-tested suites used to perform such studies are the \texttt{PyBoltz} code \cite{al2020electron}, its predecessor \texttt{MagBoltz} code \cite{biagi1999monte} and the recent \texttt{ThunderBoltz} which is a 0D Direct Simulation Monte-Carlo (DSMC) code \cite{park2024thunderboltz} also requires cross section set to perform plasma transport studies. Before delving into these studies, it is crucial to understand how the gas behaves when exposed to charged particles such as electrons and positrons. Our aim is to provide the fundamental electron scattering cross sections for some of the important gases used in high-energy physics experiments such as the electron impact total ionization cross sections (TICS) and partial ionization cross sections (PICS). In the absence of the PICS data, only the TICS is being used to perform the studies on calculating swarm parameters \cite{metting2024electron}. Hence, including the dissociative ionization channels will certainly improve the accuracy of the calculations of swarm parameters. Moreover, the data obtained from these collision processes are used to calculate the electron transport parameters using the bolsig+ \cite{hagelaar2016brief}, in a weakly ionized gas medium.  The molecules that are investigated in this article are the Trifluoromethane (CHF$_3$) also commonly known as Fluoroform is found to have zero GWP and zero ozone depletion potential (ODP), Whereas the 1,1,1,2-Tetrafluoroethane (CH$_2$FCF$_3$) commonly known as the R134a has a GWP of 1430 and zero ODP. 1,1,1-Trifluoroethane (C$_2$H$_3$F$_3$) also identified as R143a has a GWP of 4300, however, it has zero ODP. The GWP and ODP of 1,1,1-Trifluoropropane (C$_3$H$_5$F$_3$) and 3,3,3-Trifluropropene (C$_2$H$_3$CF$_3$) are unavailable in the literature. In the present article, we have calculated the TICS of the neutral molecule and PICS of the prominent cationic fragments whose appearance energies and relative abundance are available in the literature. The cross section obtained is used to calculate the corresponding rate coefficients for TICS and PICS using the Maxwell-Boltzmann distribution. The article's structure is as follows: \cref{sec2} describes the models applied to calculate the PICS; \cref{sec3} discusses the calculation of rate coefficients for ionization processes and in  \cref{sec4} we have discussed the results and findings of the present study. 
\begin{figure*}
     \centering
     \begin{subfigure}[b]{0.18\textwidth}
         \centering
         \includegraphics[width=\textwidth]{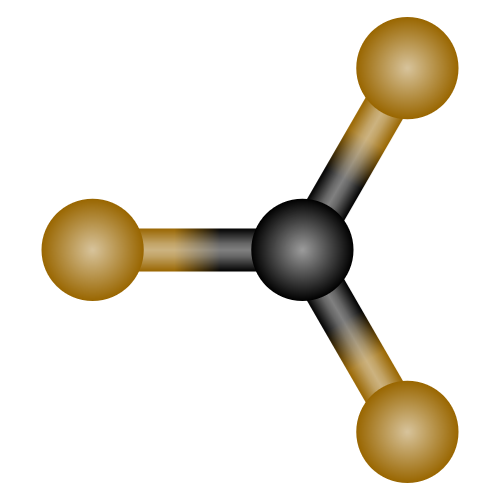}
         \caption{$\mathrm{CHF_3}$}
         \label{fig:Trifluoromethane-struct}
     \end{subfigure}
     \hfill
     \begin{subfigure}[b]{0.18\textwidth}
         \centering
         \includegraphics[width=\textwidth]{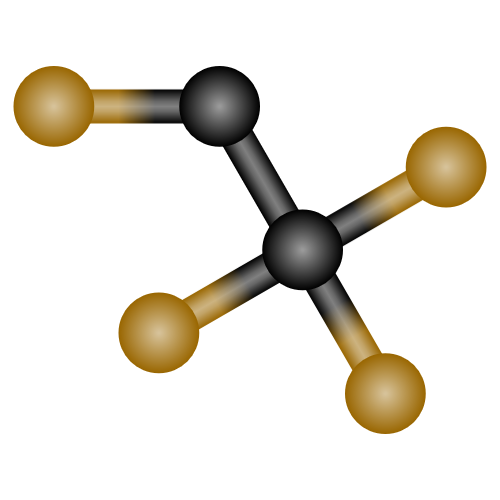}
         
         \caption{$\mathrm{CH_2FCF_3}$}
         \label{fig:1112tetrafluoroethane-Struct}
     \end{subfigure}
     \hfill
     \begin{subfigure}[b]{0.18\textwidth}
         \centering
         \includegraphics[width=\textwidth]{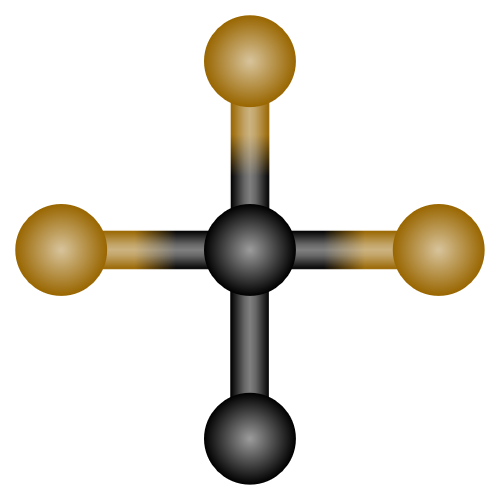}
         
         \caption{$\mathrm{CH_3CF_3}$}
         \label{fig:111Trifluoroethane-struct}
     \end{subfigure}
     \hfill
      \begin{subfigure}[b]{0.18\textwidth}
         \centering
         \includegraphics[width=\textwidth]{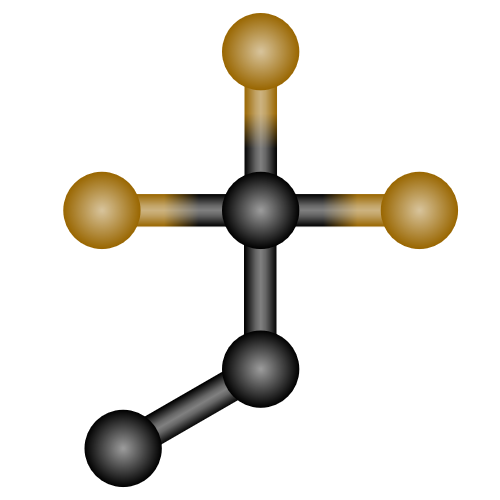}
         

            \caption{$\mathrm{C_2H_5CF_3}$}
         \label{fig:111Trifluoropropane-Struct}
     \end{subfigure}
     \hfill
     \begin{subfigure}[b]{0.18\textwidth}
         \centering
         \includegraphics[width=\textwidth]{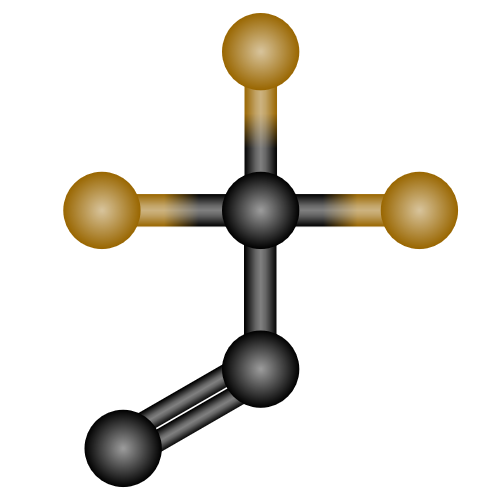}
         
         \caption{$\mathrm{C_2H_3CF_3}$}
         \label{fig:333trifluoropropene-Struct}
     \end{subfigure}
     
    \caption{The structures of the molecular targets: a) Trifluoromethane b)1,1,1,2-Tetrafluoroethane  (R134a)  c) 1,1,1-Trifluoroethane (R143a),  d) 1,1,1-Trifluoropropane, e) 3,3,3-Trifluropropene}
    \label{fig: Structures}
\end{figure*}

\section{BEB Model for electron}\label{sec2}
The binary encounter Bethe (BEB) model is a widely used formalism to calculate the electron and positron impact ionization cross sections of atoms, molecules, and ions. Many works \cite{gupta2017electron,shanmugasundaram2024electron} have employed the BEB model to predict the TICS and PICS by electron impact of the molecular targets. The model generally works by calculating the ionization cross sections of each occupied orbital present in the molecule and then summing it up to provide the TICS for electron impact as given in \Eref{eq:1}. 
\begin{equation}\label{eq:1}
    \sigma_{TICS}(E) = \sum_{i}^{N} \sigma_{i}(E)
\end{equation}
The simplified BEB model introduced by Kim and Rudd \cite{kim1994binary} is defined as,

\begin{eqnarray}\label{eq:2}
 \sigma_i^{BEB}(E) &= {S_i\over  (t_i+u_i+1)/ n}  \Bigg[{ Q_{i} \ln t_i\over  2}\left ( 1 - {1\over  t_i^{2}}\right) \nonumber \\  
 & + (2-Q_{i}) \left \{ \left ( 1 - {1\over  t_{i}} \right )- {\ln t_{i}\over  t_{i}+1} \right \} \Bigg]
\end{eqnarray}
Here, $t_{i},u_{i}$ and S are the reduced variables which are defined as,
\begin{equation}\label{eq:3}
S_i = 4\pi a_0^2 N \left ({R\over B}\right )^2,~u_i={U\over B},~t_i={E\over B}
\end{equation}
In the present calculation, $a_0$ and R represent the Bohr radius (0.52918 $\AA$), and the Rydberg constant (13.6057 eV). The differential oscillator strength, $Q_i$ is set to unity according to the simplified BEB model. The U and B are the orbital kinetic and binding energy of the molecule which is obtained by performing quantum chemistry calculations. The scaling factor $(n)$ is fixed to unity for all the molecular orbitals due to the presence of lighter atoms $(Z<10)$ in the molecular targets. If there exists an atom with $(Z>10)$, $(n)$ is set as the principal quantum number of the dominant atom, which is determined by performing Mulliken population analysis (\cite{shanmugasundaram2024electron} and references within). N is the occupation number of the orbital, and the incident electron kinetic energy is represented by E. The orbital parameters for a molecule were calculated using a similar method applied in our previous work \cite{shanmugasundaram2024electron}, where we have optimized the geometries of the molecule by applying the density functional theory (DFT) with the functional ($\omega$B97XD) and with the aug-cc-PVTZ (aVTZ) basis set. Once the stable geometry is obtained then the energies are calculated using the Hartree–Fock ($\mathcal{HF}$)
approximation using the same basis set (aVTZ). In \Fref{fig: Structures}, the structure of the molecular targets sketched from the SMILES code from the Protein Data Bank's online chem sketch tool \cite{rcsbRCSBPDB} is presented. The $\mathcal{HF}$ method generally overestimates the binding energies a little which leads to the underestimation of the theoretical TICS when compared with the experimental cross sections. However, the calculated TICS is found to lie between 10 \% and 15 \% uncertainty when compared with the experiment \cite{gupta2017electron}. 
\subsection{mBEB model}
To calculate the PICS we need information such as the fragmentation pathway, dissociation energy/appearance energy $(\varphi)$, and the electron impact mass spectrum (EIMS). We will be discussing two methods to calculate the PICS, the mass spectrum dependence (MSD) method and the modified BEB (mBEB) method. In the mBEB model, the cations formed due to dissociative ionization by electron or positron impact can be distinguished through $\varphi$. The ionization energy of the molecule is obtained from the highest occupied molecular orbital (HOMO) of the neutral parent molecule ( IP$_{koopman's}$ = - HOMO ). To enable the mBEB model to calculate the PICS for all cations using AE and experimental BR, a minor adjustment to the binding energies of the occupied orbitals of the neutral parent molecule is required. This is facilitated via summing the difference $(\Delta)$ between the neutral parent molecule's IE and the $(\varphi)$ of the fragment, $(\Delta = IE-\varphi)$ to all the values of the orbital binding energies of the parent molecule $(B'=B+\Delta)$. This correction replaces the HOMO of the cation by its $\varphi$.
The mBEB method is defined as,
\begin{eqnarray}\label{eq:4}
\centering
\sigma_i^{mBEB}(E)&={S'_i\over  (t'_i+u'_i+1)/ n}\Bigg [ { Q_{i} \ln t'_i\over  2}\left ( 1 - {1\over  t_i^{'2}}\right) \nonumber \\
&+ (2-Q_{i}) \left \{ \left ( 1 - {1\over  t'_{i}} \right )- {\ln t'_{i}\over  t'_{i}+1} \right \} \Bigg ]
\end{eqnarray}
The modified reduced variables are as follows,
\begin{equation}\label{eq:5}
    S'_i = 4\pi a_0^2 N \left ({R\over B'}\right )^2,~t'_{i} = \left ( {E\over B'} \right ),~u'_i={U\over B'}
\end{equation}
Then the PICS is calculated as shown in \Eref{eq:6},  
\begin{equation}\label{eq:6}
    \sigma^{PICS} (E) = \Upsilon_i (E_r) \times \sigma_i^{mBEB}(E)
\end{equation}
$\Upsilon_i$  is the scaling factor which is determined from the ratio of the experimental BR vs the theoretical BR [see  \Eref{eq:09}]. The BR obtained via the experimental EIMS data is the experimental BR $(\Gamma_i^{Exp})$ which is estimated as seen in  \Eref{eq:7}. In experiments, the EIMS data is measured at one incident energy or reference energy ($E_r$) which can either be at 70 eV, 100 eV or 125 eV according to the user's choice of the experimental parameter,
\begin{equation}\label{eq:7}
\centering
    \Gamma_i^{Exp}(E_{r}) = {\rm R(E_{r})\over  T(E_{r})}
\end{equation} 
Here the R(E$_{r}$) is the relative abundance of the cation and the T(E$_{r}$) is the total abundance, the sum of all R(E$_{r}$) of cations. This electron impact mass spectrum can also be generated theoretically using quantum chemical mass spectrometry (QCxMS)\cite{koopman2021qceims}, when the experimental EIMS data are scarce. 
The theoretical BR $(\Gamma_i^{Theo})$ is the ratio of the PICS calculated with the mBEB model before introducing the scaling factor,
\begin{equation}\label{eq:08}
        \Gamma_i^{Theo}(E_r)  = {\sigma_i^{mBEB}(E_r)\over  \sigma_i^{BEB}(E_r)}
\end{equation} 
Then the scaling factor $\Upsilon_i (E_r)$ is obtained by
\begin{equation}\label{eq:09}
\Upsilon_i (E_r)  = {\Gamma_i^{Exp}(E_r)\over  \Gamma_i^{Theo}(E_r)}
\end{equation}
There is also another method that was implemented by Huber et al. \cite{huber2019total} which can be used to calculate the BR using the appearance energy or the dissociation energy which we have explained in our recent work \cite{shanmugasundaram2024electron}. The MSD method to compute PICS is discussed in the next section.
\subsection{MSD method}
Usually, the branching ratio is a single-valued quantity in the above mBEB method. But, here we make it incident energy-dependent and thus continuous throughout the range of the incident kinetic energy (E). Here we also need the information on the experimental BR. 
\begin{equation}\label{eq:10}
 \small{\Gamma_i^{MSD} (E)=  \cases{ 
0 & if E $< \varphi$  \\ 
 \Gamma_i(E_r)\Big [ 1 - \Big ( {\varphi \over E} \Big )^{\nu} \Big ] & if E$\geq \varphi$ \\ }}
\end{equation}
The value $\nu$ is set to be $1.5$, which is the control parameter suggested by Janev et al. \cite{janev2004collision} More details about this method can be obtained from the works of Graves \cite{graves2022calculated}, Huber\cite{huber2019total}, and in our recent work\cite{shanmugasundaram2024electron}. Another important component of the BR is that its sum always yields unity, irrespective of the method used. In a few cases where we might not know all the fragmentation pathways of a molecule, the sum can be less than unity. Once we perform this BR check, we can calculate the PICS using the MSD method.
\begin{equation}\label{eq:11}
    \sigma^{PICS} (E) = \Gamma^{MSD}_i (E) \times \sigma_i^{BEB}(E)
\end{equation}

\section{Ionization Rate coefficients}\label{sec3}
In thermal plasmas, the electron's velocity distribution function is considered as the Maxwellian electron distribution function for isotropic velocities for temperature. 
\begin{equation}\label{eqn:20}
    f(v,T_e) = 4\pi v^2\Bigg( {m_e \over 2\pi k_BT_e}\Bigg)^{3/2} \exp \Bigg(-{m_e v^2 \over 2k_BT_e} \Bigg)
\end{equation}
Where,  $v$ is the electron velocity, $m_e$ is the mass of the electron and $k_B$ is the Boltzmann constant. The rate coefficient is derived by averaging the cross sections over the Maxwellian electron velocity distribution (MVD) as follows.
\begin{equation}\label{eqn:21}
    k(T_e) = \int _0^\infty \sigma(v)f(v,T_e) vdv
\end{equation}
$\sigma(v)$ is the ionization cross sections. The calculated ionization rate coefficients are in the temperature range of 1 eV and 10 eV which is then fitted using the Arrhenius equation till 100 eV temperature, 
\begin{equation}\label{eqn:22}
    k = A T^n_e \exp \Bigg( {-E_{act}\over T_e}\Bigg)
\end{equation}
$ A T^n_e$ is the temperature dependent pre-factor in $\rm{cm^3s^{-1}}$, $ E_{act}$ is the activation energy in eV and $T_e$ is the electron temperature in eV. $ E_{act}$, A, $n$ are the fitting parameters. The best-fit parameters are given in the \cref{tab:fit-para,tab:fluoroform,tab:tfe,tab:111tfp,tab:333tfp,tab:1112tfe}.


\section{Results and Discussion}\label{sec4}
In this part, we show the PICS and TICS calculated for all the molecular targets along with their corresponding rate coefficients. Since the BEB cross section is sensitive to the values of the valence shell orbitals, the HOMO energies obtained in our calculations using the $\mathcal{HF}$ and DFT approach are presented along with the literature values in \cref{tab1-IE}. Each sub-section discussed below contains the branching ratio, PICS, TICS, and the corresponding rate coefficient for PICS and TICS for each of the targets studied. 

\begin{table*}
\centering
\caption{Vertical, Adiabatic, HOMO, Ionization energies of molecules}
\label{tab1-IE}
\begin{tabular}{lllll}
    \br
    Molecular  & Name & \multicolumn{2}{c}{HOMO (eV)} & Ionization potential (eV) \\
   formula& & DFT & $\mathcal{HF}$  & Literature \\
    \mr
    $\rm{CHF_3}$ & Trifluoromethane & 13.40 & 16.31  &13.86\cite{lias_ionization_2024}, 14.8\cite{harshbarger_electron_1973},15.5$^b$\cite{linstrom2001nist}\\
    $\rm{CH_3CF_3}$ & 1,1,1 Trifluoroethane & 12.58 & 15.39  & 13.3 $\pm$ 0.1 \cite{simmie1971mass}, 13.8$^b$\cite{linstrom2001nist} \\
    $\rm{C_2H_5CF_3}$ & 1,1,1 Trifluoropropane &10.69 & 11.42 & $-$  \\
    $\rm{C_2H_3CF_3}$ & 3,3,3 Trifluoropropane&  12.16 &  14.22 & 11.24\cite{steele1962electron}  \\ 
    &~~&~~&~~& \\
    \multirow{2}{*}{$\rm CH_2FCF_3$} & \multirow{2}{*}{1,1,1,2 Tetrafluoroethane}& \multirow{2}{*}{12.48} & \multirow{2}{*}{15.16} & 13.10 $\pm$ 0.17\cite{pereira2021electron}, 13.19\cite{pereira2021electron}, \\
     &&  &  & $12.25^{a}\cite{zhou2002fragmentation} 13.96^{b}$\cite{zhou2002fragmentation}   \\
    \br 
    \small{$^{a}$Adiabatic IE,$^{b}$Vertical IE}
\end{tabular}
\end{table*}



\subsection{Trifluoromethane}\label{sec:fluoroform}
 CHF$_3$ molecule has been very well studied over several years for ionization experimentally. Here we would like to revisit them and also theoretically calculate the PICS as they are not available in the literature and the corresponding rate coefficients for ionization and dissociative ionization are also computed. Goto and co-workers \cite{goto1994cross} had experimentally measured the TICS and PICS of CHF$_3$ using the dual electron beam apparatus coupled with a quadrupole mass spectrometer (QMS). They provided the reaction channels for neutral dissociation and also measured the threshold energy. It was the first time where the cross sections were reported for individual cations along with their AEs, until then only the dominant cations CF$_x^+: x=1,2,3$ and combined cross sections of two or more cations were reported \cite{hobrock1964electron,sugai1995absolute}. Later, Jiao et al \cite{jiao1997ion} also provided the experimental PICS and TICS for the CF$_x^+: x=1,2,3$ along with CHF$_2^+$ using the Fourier transform mass spectrometry (FTMS) with a cubic ion cyclotron resonance (ICR) trap cell. Iga and colleagues \cite{iga2001electron} also provided the PICS, TICS, and the mass spectrum for the same. The measurements were performed using the time-of-flight mass spectrometer (TOFMS) along with the QMS for identification of $m/z$ values of ions, the mass spectrum was recorded at 200 eV incident energy. Their mass spectrum shows that the base peak was from CHF$_2^+$ ion and the other peaks are due to ions of CF$_x^+ ( x=1,2,3)$, C$^+$ and F$^+$. The PICS of these cations detected in the mass spectrums were measured from the ionization threshold till 1 keV, and cross sections of CF$_3^+$ and F$^+$ ions were measured individually. The other contributions of PICS are bundled cross sections of the ions (CHF$_2^+$ + CF$_2^+$), (CHF$^+$ + CF$^+$) and the (C$^+$ + CH$^+$), their findings agree very well with Goto et al \cite{goto1994cross}. Torres et al \cite{torres2002electron} experimentally measured the PICS using the linear double-focusing time-of-flight apparatus where the electron kinetic energies were varied from 0 eV to 100 eV, along with their AEs. 

On the theoretical front, Onthong\cite{onthong2002calculated} working with Deutsch and M\"ark used the Deutsch–M\"ark (DM) model to calculate the TICS, their data had a very good comparison with the Iga et al \cite{iga2001electron}. Torres et al \cite{torres2001evaluation} has extensively performed TICS calculations using parameters from various levels of theory such as the M$\o$ller-Plesset second-order perturbation (MP2), M$\o$ller-Plesset fourth-order perturbation (MP4), Configuration interaction singles and doubles (CISD), Coupled cluster (CC) and $\mathcal{HF} $ for several fluoromethanes including CHF$_3$. They replaced the HOMO energy with an experimental ionization threshold value for better results from the BEB method. In their work, they also compared the cross-sections calculated using the modified additivity rule (MAR), DM formalism, and the BEB method. It is well known that the MAR would overestimate the TICS and the DM method also falls short in low energy regions. The calculation using the BEB model included the optical oscillator strengths (OOS) of the ground state of the Hydrogen atom as an approximation rather than using the OOS of target molecules. The OOS are quite cumbersome to calculate. In our case, we have used the simplified BEB model which does not require the OOS. 

\begin{figure*}[h]
    \begin{tikzpicture}
    \node at (0,0) {\includegraphics[width=\textwidth]{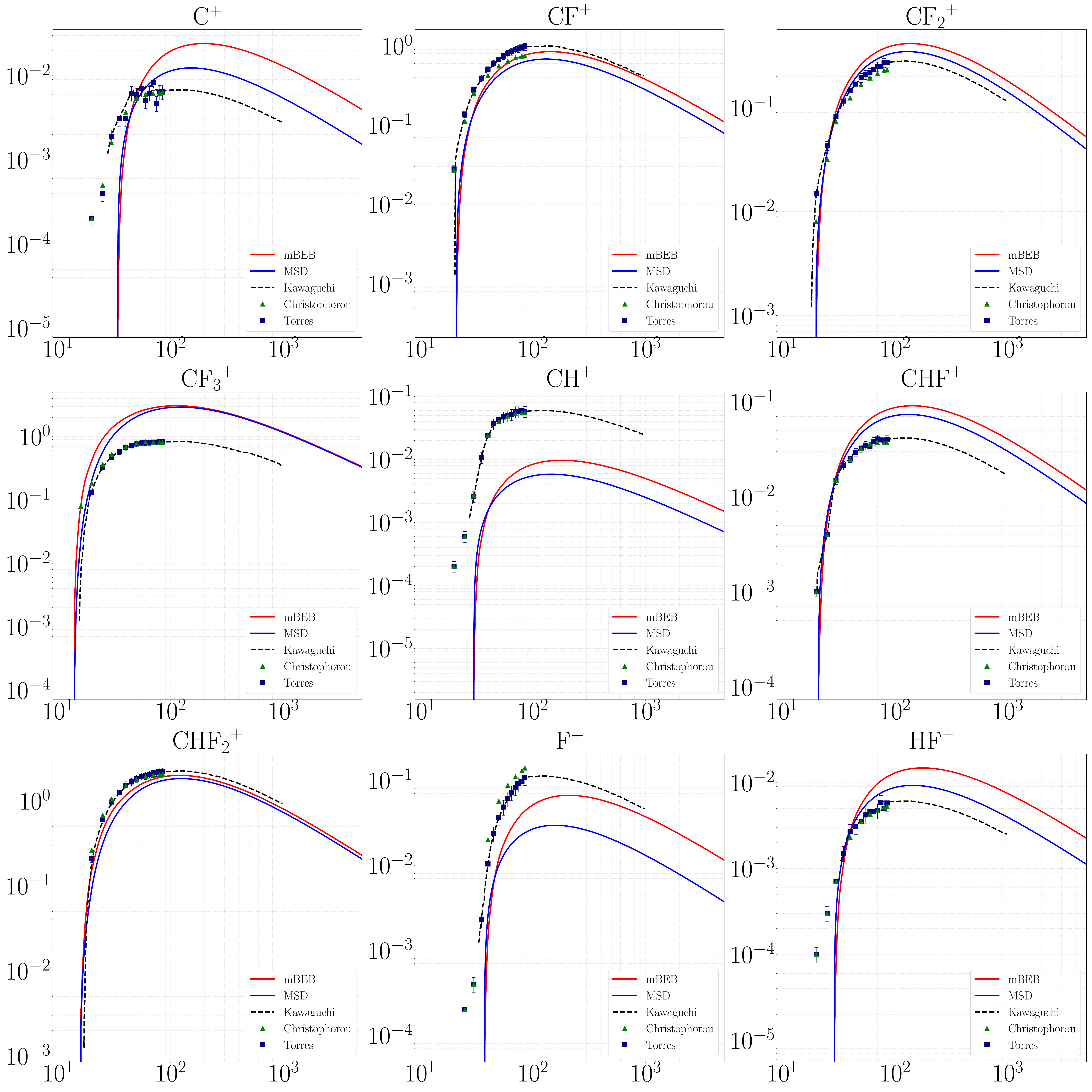}};
    \node[below, yshift=-0.5ex] at (current bounding box.south) {Energy (eV)};
    \node[left, rotate=90, xshift=9.5ex] at (current bounding box.west) {PICS $(10^{-16}cm^{2})$};
  \end{tikzpicture}
    \caption{PICS of Trifluoromethane:  cations detected in the EIMS, the blue solid line represents the calculated cross sections using the mBEB model, and the solid red line represents the calculated cross sections using the MSD method. the dotted lines represent the PICS presented by Kawaguchi et al\cite{kawaguchi2014electron}, the green upright triangles represent the recommended cross sections by Christophorou et al \cite{christophorou2004electron} and the blue squares are the PICS by Torres et al \cite{torres2002electron}.}
    \label{fig:CHF3-pics}
\end{figure*}

\begin{figure*}
\begin{subfigure}[t]{0.5\textwidth}
    \centering
    \includegraphics[width=\textwidth]{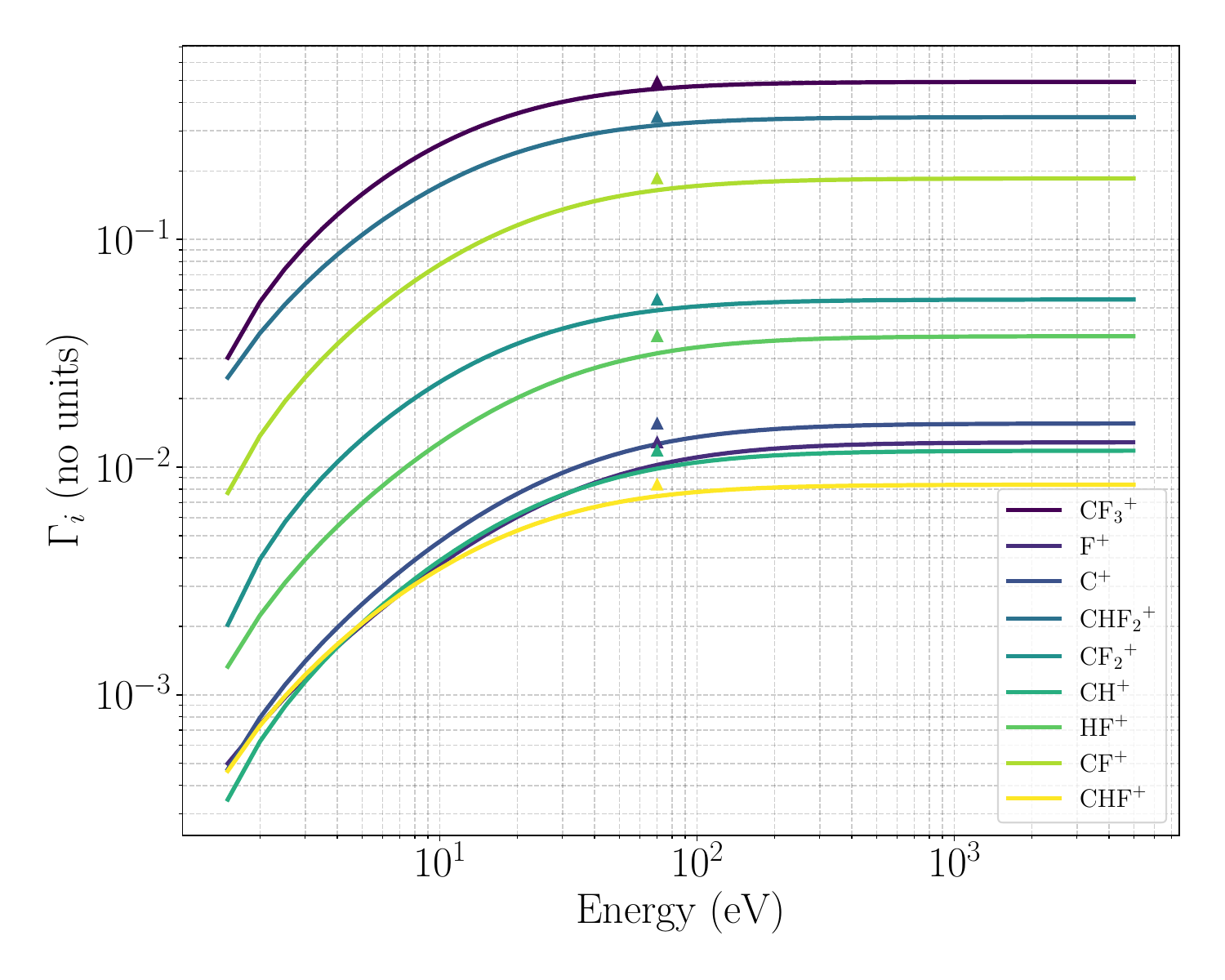}
    \caption{}
    \label{fig:CHF3br}
\end{subfigure}
\begin{subfigure}[t]{0.5\textwidth}
    \centering
    \includegraphics[width=\textwidth]{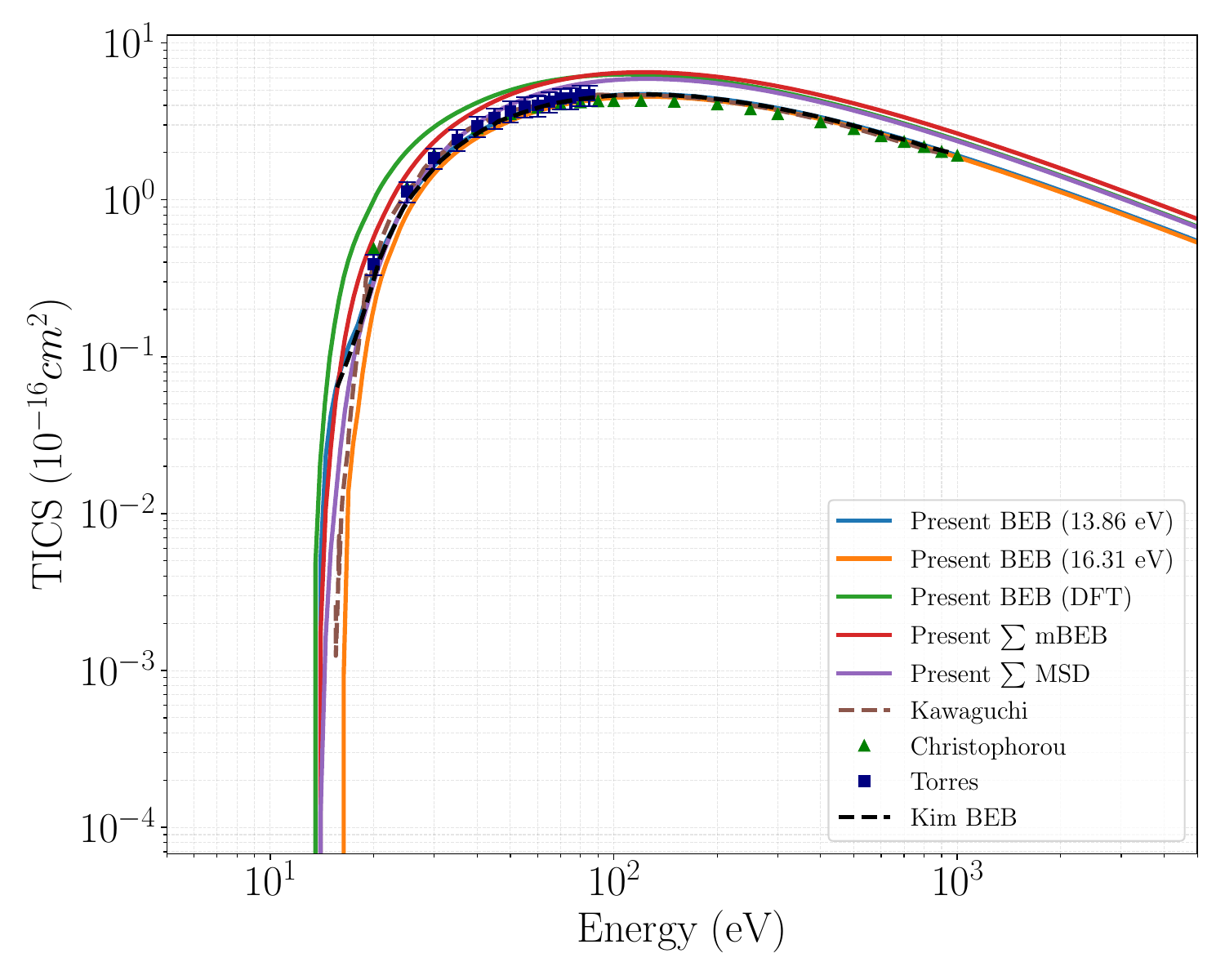}
     \caption{}
    \label{fig:CHF3tics}
\end{subfigure}
\caption{a) Branching ratios for the fragments of Trifluoromethane: The solid lines represent the branching ratios calculated for the MSD method, whereas the solid triangles denote the branching ratios calculated from the EIMS. b) Trifluoromethane TICS calculated using different methods, the solid blue line represents our calculated BEB TICS where the experimental first ionization energy was used, and the solid orange line shows the present BEB TICS calculated using HOMO energy obtained by HF approximation. The solid green line represents the TICS calculated from orbital parameters obtained using DFT. The red solid line shows the TICS which was the sum of all the PICS calculated from the mBEB model. 
The solid purple line shows the TICS which was the sum of all the PICS calculated from the MSD model. The dashed brown line represents the TICS from the work of Kawaguchi et al \cite{kawaguchi2014electron}. The solid upright triangles represent the recommended TICS from Christophorou et al \cite{christophorou2004electron}. The dashed black line represents the TICS calculated using the BEB method by Kim et al \cite{torres2001evaluation}. }
\end{figure*}

In our present work, we have considered the AEs from the work of Torres et al \cite{torres2002electron} for each cation which can be seen in \Tref{tab:fluoroform}. We would also require the mass spectrum for calculating the BR. The electron impact mass spectrum data is provided by Peko et al \cite{peko2002measured} measured at the incident energy 70 eV, where they have detected most of the fragments but have missed detecting the HF$^+$ fragment. In our calculations, we have chosen the EIMS data from the NIST Webbook \cite{linstrom2001nist}, for calculating the BRs and shown in \Tref{tab:fluoroform}.  We have presented the BR calculated using the MSD method and the BR calculated from the experimental branching ratios in  \Fref{fig:CHF3br}. Later Christophorou and Olthoff \cite{christophorou2004electron} provided the recommended TICS and PICS by compiling the PICS of Jiao et al \cite{jiao1997ion}, Iga et al \cite{iga2001electron} and Torres et al \cite{torres2002electron} altogether by fitting, summing and normalizing the cross sections for each energy. It can be seen in \Fref{fig:CHF3-pics}, that the recommended cross sections and our calculations match very well for most stable ions. It was also mentioned by Torres et al that the uncertainties for the lighter ions $(m/z < 20 ~\rm amu )$ are higher (15\% to 20\%) whereas the heavier ions tend to have an uncertainty of less than 10\%. The recommended data by Christoporou et al \cite{christophorou2004electron} does not have any uncertainties. 
Recently, Kawaguchi et al. \cite{kawaguchi2014electron} used the inverse swarm method to calculate the PICS from 85 eV to 1 keV using the TICS presented by Kim et al.

\begin{figure*}
    \begin{tikzpicture}
    \node at (0,0) {\includegraphics[width=\textwidth]{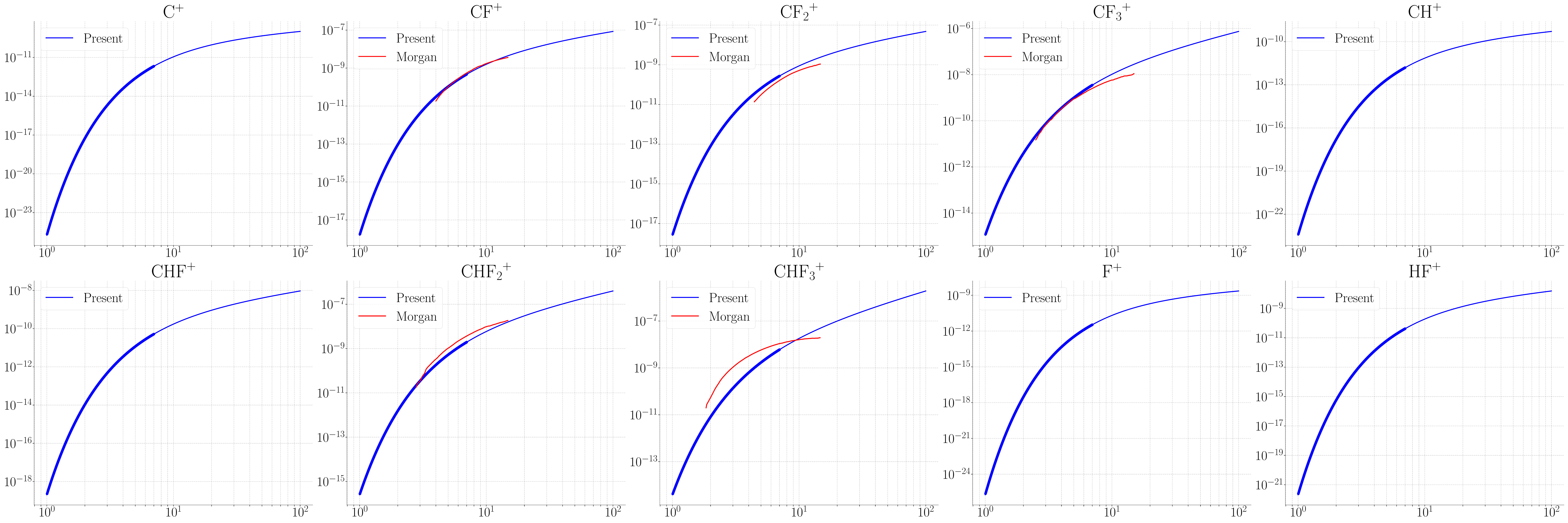}};
    \node[below, yshift=-0.5ex] at (current bounding box.south) {Temperature (eV)};
    \node[left, rotate=90, xshift=9.5ex] at (current bounding box.west) { $k (cm^{3}s^{-1})$};
  \end{tikzpicture}
    \caption{Ionization rates of Trifluoromethane: the blue solid lines represent the ionization rates, whereas the red solid lines represent the rates calculated presented by Morgan et al \cite{morgan2001electron} in the range of 1 eV till 15 eV.}
    \label{fig:CHF3-rates}
\end{figure*}

From \Fref{fig:CHF3-pics}, it can be seen that for lighter fragments such as the CH$^+$ and F$^+$, there is quite a lot of disagreement of our cross sections when compared with the literature data. For all the other fragments our theoretical cross sections are similar to the literature values. We have calculated the ionization and dissociation rates in \Fref{fig:CHF3-rates}, where we have used the cross sections calculated with the MSD method as an input since it had a better agreement of cross sections with the literature data as shown in our previous work \cite{shanmugasundaram2024electron} as well as in the current work. Morgan-Winstead-McKoy\cite{morgan2001electron} calculated the dissociation rates for the fragments CF$_x^+: x=1,2,3$ and CHF$_y^+: y=2,3$ in the range of 0 eV and 15 eV. Later Bose and co-workers calculated the rate coefficients and fitted them with the Arrhenius equation, provided with the necessary parameters. In \Tref{tab:fit-para}, we have compared the fitting parameters calculated using Scipy's curve fit function \cite{scipy} with the literature values.
\begin{table*}[h]
\caption{The calculated fitting parameters of the Arrhenius equation to calculate rate constants for CHF$_3$ molecule compared with the data shown by Bose et al \cite{bose2003uncertainty}}
\begin{tabular}{llccccc}
\br 
 \multirow{2}{*}{Cation}& \multicolumn{2}{c}{A ($10^{-10}$cm$^3$s$^{-1}$ eV$^{-n}$)} & \multicolumn{2}{c}{$n$} & \multicolumn{2}{c}{$\mathrm{E_{act}}$ (eV)} \\ 
&Present&Literature&     Present      &     Literature     &   Present        &    Literature      \\
 \mr
 CF$_3$&22.8&    313.8\cite{bose2003uncertainty}      &   1.288        &     -0.2366\cite{bose2003uncertainty}   &      14.5532     &     $-$        \\
\multirow{2}{*}{CHF$_2$}&      \multirow{2}{*}{17.4}    &   184.5\cite{morgan2001electron},       &       \multirow{2}{*}{1.2142}    &    0.4022\cite{morgan2001electron},       &   \multirow{2}{*}{15.7526}        &  17.08\cite{morgan2001electron},    \\
 &           &    48.59\cite{bose2003uncertainty}       &        &     -0.1043\cite{bose2003uncertainty}      &           &   19.576\cite{bose2003uncertainty}    \\
 \multirow{2}{*}{CF$_2$}&     \multirow{2}{*}{6.37}      &  242.9\cite{morgan2001electron},     &      \multirow{2}{*}{0.9797}     &      0.2566\cite{morgan2001electron},    &       \multirow{2}{*}{19.3258}    & 22.33\cite{morgan2001electron}       \\
 &          &  33.29\cite{bose2003uncertainty}     &           &       -0.0048\cite{bose2003uncertainty}    &         &  19.542\cite{bose2003uncertainty}         \\
 CHF&      1.56     &    20.79\cite{bose2003uncertainty}      &    0.9327       &   0.4581\cite{bose2003uncertainty}       &      20.4678     &     $-$      \\
\multirow{2}{*}{CF}&     \multirow{2}{*}{1.49}     &       119.0\cite{morgan2001electron},   &  \multirow{2}{*}{0.9224}         &   0.1276\cite{morgan2001electron},       &       \multirow{2}{*}{20.6763}   &  23.16\cite{morgan2001electron},         \\
 &          &        64.37\cite{bose2003uncertainty}   &         &   0.2857\cite{bose2003uncertainty}       &         &  19.474 \cite{bose2003uncertainty}        \\
 HF&     5.2      &    $-$      &   0.7968        &      $-$       &    28.55       &       $-$      \\
 F&        2.04   &      $-$       &     0.6069      &    $-$         &   36.9257        &        $-$     \\
 CH&         0.208  &       $-$      & 0.6955          &     $-$        &   29.7174        &         $-$    \\
 C&         0.723  &        $-$     &    0.6484       &       $-$      &      33.6389     &           $-$  \\
 CHF$_3$&     21.1      &     103.3\cite{morgan2001electron}     &    1.5084       &       0.3601\cite{morgan2001electron}     &  13.2155         &     15.00\cite{morgan2001electron}    \\
 \br 
\end{tabular}
\label{tab:fit-para}
\end{table*}
\begin{table*} [h]
\caption{ The Appearance Energy (AE), Branching ratios (BR), cross section's maximum value ($\sigma$), the values of pre-exponent (A), scaling term $(n)$ and the activation energy $(\rm E_{act})$ of Trifluoromethane (CHF$_3$) }
\label{tab:fluoroform}
\begin{tabular*}{\textwidth}{@{}l*{15}{@{\extracolsep{0pt plus
12pt}}l}}
\br
$m/z$&Cation& AE (eV)& BR&$ \sigma$ (10$^{-16}$cm$^2$)&A($10^{-10}$cm$^3$s$^{-1}$ eV$^{-n}$)&$n$ & E$_{act}$ (eV)\\
\mr
69&CF$_3$&13.9 $\pm$ 0.6 &0.49 263&2.2397&5.693 &1.7021
&14.3532\\
51&CHF$_2$&15.7 $\pm$ 0.5 &0.34 464&1.5561&6.183&1.5269
&15.8342\\
50&CF$_2$&19.5 $\pm$ 0.5 &0.05 449&0.2423&2.317&1.5269
&19.3377\\
32&CHF&20.7 $\pm$ 0.5&0.00 838&0.0371&4.368&1.1217
&20.4676\\
31&CF&20.9 $\pm$ 0.5&0.18 548&0.8198&10.063&1.1072
&20.6735\\
20&HF&28.8 $\pm$ 0.5&0.06 644&0.1605&5.196&0.7968
&28.55\\
19&F&37.0 $\pm$ 0.5&0.01 286&0.0528&2.917&0.6547
&36.8771\\
13&CH&29.9 $\pm$ 0.5&0.01 181&0.0502&1.786&0.7693
&29.6736\\
12&C&33.8 $\pm$ 0.5&0.01 555&0.0649&2.968&0.7035
&33.5933 \\ 
70&\textbf{CHF$_3$}&\textbf{13.860}&$-$&4.7150&21.108&1.5084
&13.2155 \\
\br

\end{tabular*}
\end{table*}

\subsection{1,1,1,2-Tetrafluoroethane}

\begin{figure*}[h]
\begin{subfigure}[t]{0.5\textwidth}
    \centering
    \includegraphics[width=\textwidth]{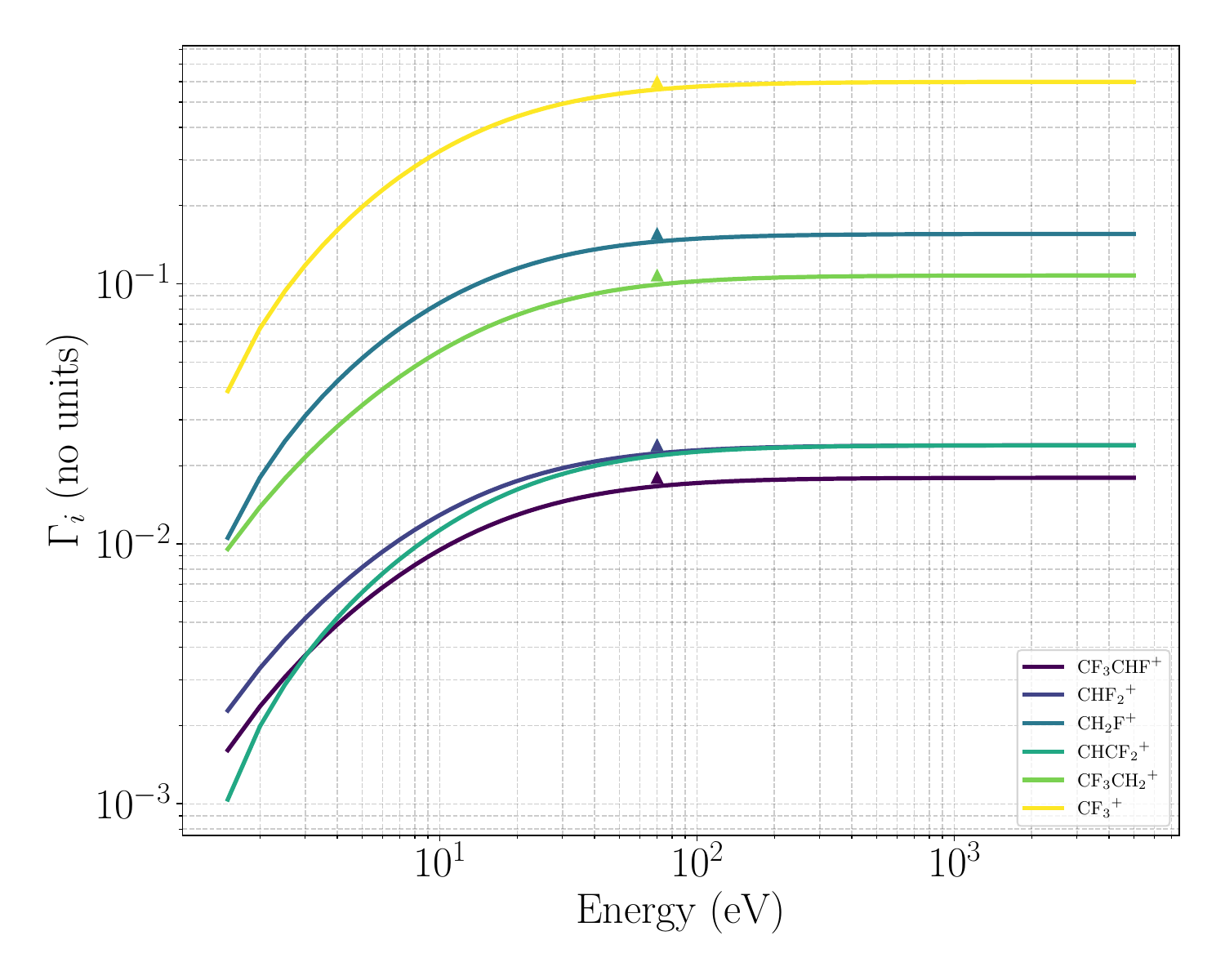}
    \caption{}
    \label{fig:r134a-br}
\end{subfigure}
\hfill
\begin{subfigure}[t]{0.5\textwidth}
\centering
    \includegraphics[height = 0.78\textwidth, width=\textwidth]{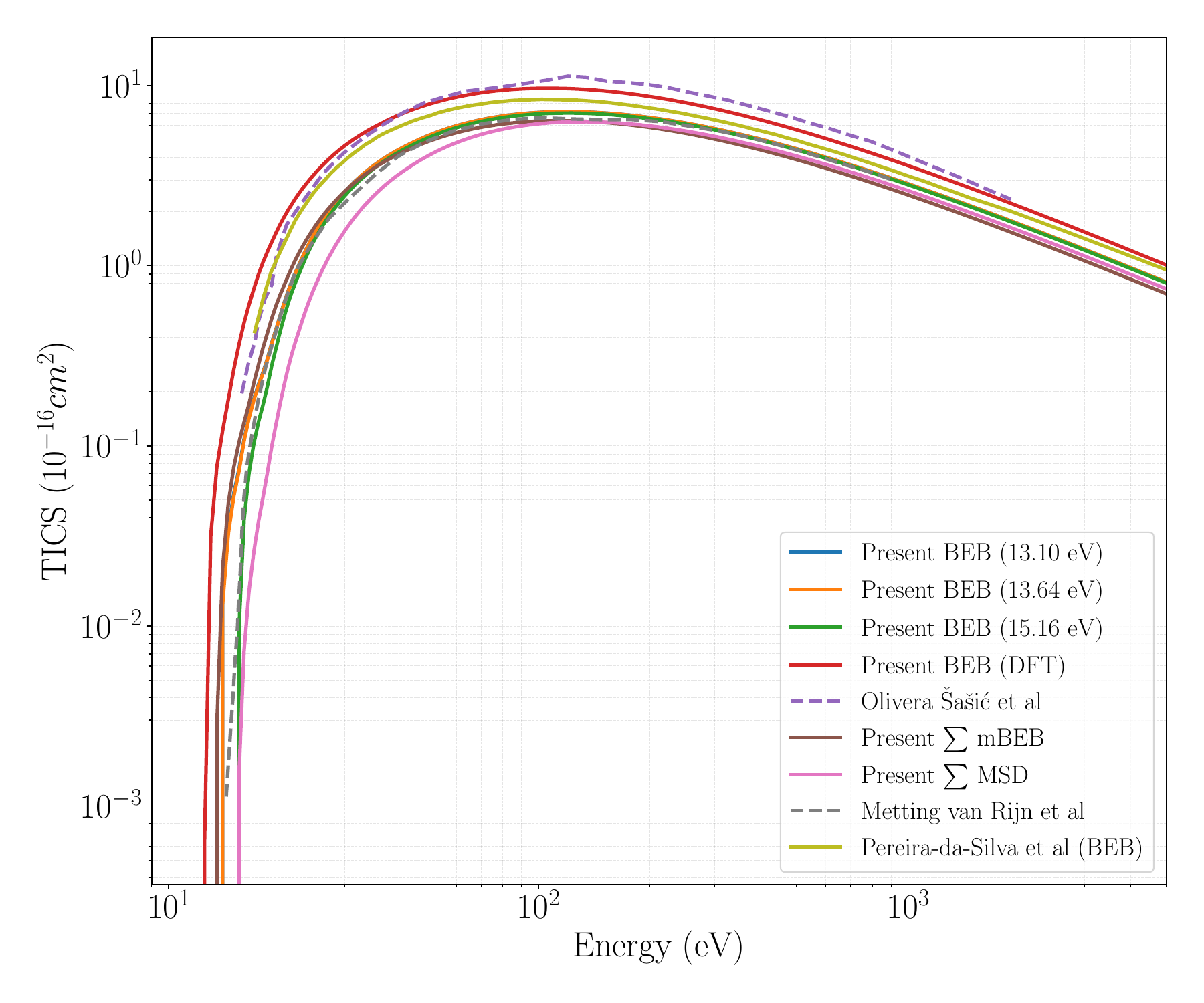}
    \caption{}
    \label{fig:TICS of 1,1,1,2-Tetrafluoroethane}
\end{subfigure}
\hfill
\begin{subfigure}[b]{0.5\textwidth}
\centering
    \includegraphics[width=\textwidth]{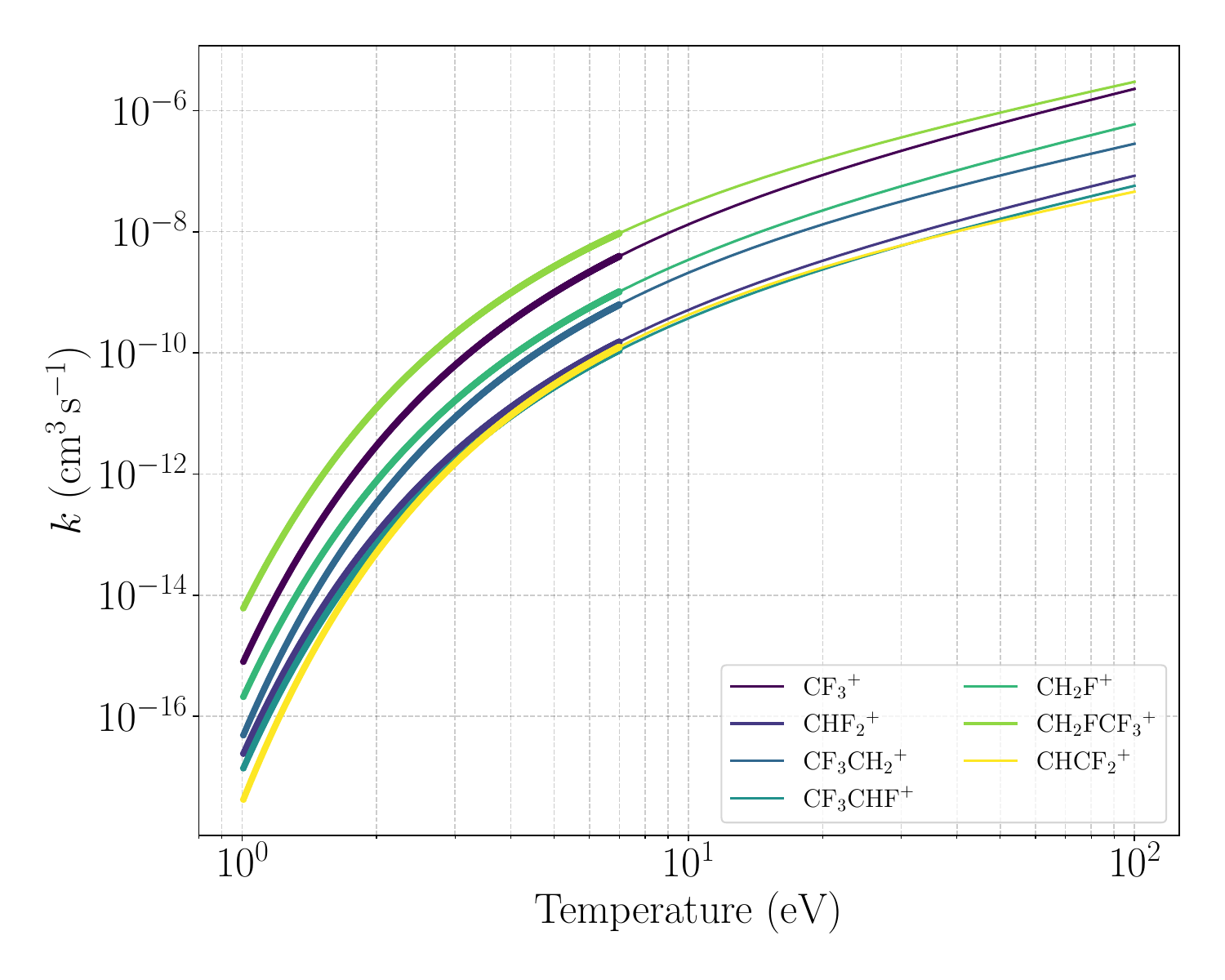}
    \caption{}
    \label{fig:r134a-rates}
\end{subfigure}
\caption{a) Branching ratios of 1,1,1,2-Tetrafluoroethane $\rm CH_2FCF_3$: The solid lines represent the branching ratios calculated for the MSD method, whereas the solid triangles denote the branching ratios calculated from the EIMS. b) TICS of 1,1,1,2-Tetrafluoroethane: The solid blue lines represent the TICS where the HOMO energy is 13.64 eV, the solid orange lines denote the TICS where the HOMO energy is set to 15.16 eV, the solid green line shows the TICS calculated using the DFT parameters, The solid red line indicates the sum of all PICS from the mBEB method, the solid purple line represents the sum of all PICS calculated from the MSD method. The dashed brown line represents the TICS presented by Marnik Metting van Rijn et al \cite{metting2024electron}, and the solid pink line represents the TICS calculated by Pereira-da-Silva et al \cite{pereira2021electron}. c) 1,1,1,2-tetrafluoroethane rate coefficients for
ionization and dissociation processes, here the bold lines are the rates calculated using the MVD and the thin solid lines are the extrapolated values calculated using the Arrhenius function from 1 eV to 100 eV. }
\end{figure*}

Here we discuss about the molecule 1,1,1,2-Tetrafluoroethane which is a quite well-studied target with the common name R134a. A plethora of data is found in the literature. The recent one is by Marnik Metting van Rijn et al has revisited the electron scattering cross section of R134a and performed swarm studies on the Pulsed Townsend apparatus experimentally and also used MAGBOLTZ to calculate parameters such as drift velocity and effective ionization co-efficient \cite{metting2024electron}, a comprehensive set of cross sections of other molecular processes such as the elastic + rotational + momentum transfer, total elastic, vibrational and the ionization has been presented. Víctor S. A. Bonfim et al investigated the photodissociation and the several pathways in the production of CHF$_2^+$ fragment (references within \cite{bonfim2024deciphering}), which had an RI of 4\% in the study of Pereira-da Silva \cite{pereira2021electron} which used a Trochoidal electron monochromator(TEMs) coupled with an Orthogonal Reﬂectron Time-of-Flight Mass Spectrometer (OReToFMS) to determine the positive and negative ion formations. They also provided the mass spectrum and AEs for a few fragments which can be seen in \Tref{tab:1112tfe}. As usual, the most abundant fragment is CF$_3^+$ having AE as 13.39 eV. Several works  \cite{pereira2021electron,hayashi2018dissociative} have cited the vertical IE as 13.2 eV which was attributed to the NIST web book \cite{linstrom2001nist}. The vertical IE and the adiabatic IE were reported by Zhou et al \cite{zhou2002fragmentation} as 12.25 eV and 13.96 eV. Here we have calculated the PICS based on the mass spectrum and AEs presented by Pereira-da Silva \cite{pereira2021electron} which is shown in \Fref{fig:PICS of 1,1,1,2-Tetrafluoroethane}. The \Fref{fig:r134a-br}, shows the branching ratios, the solid lines represent the continuous BRs and the upright triangles represent the experimental BR. The fragments CHCF$_2^+$ (63) and CHF$_2^+$(51) have the same RI which is the reason for the overlapping of the continuous BR in the high energy region. In \Tref{tab:1112tfe}, we have shown the BRs, cross sections, and the fitting parameters for the rate coefficients.
The rate constants for the TICS and PICS have been presented in \Fref{fig:r134a-rates}. The TICS has been calculated using the parameters obtained from the $\mathcal{HF}$ and DFT parameters. The HOMO values have been replaced with the literature values and the TICS profiles using various methods along with the literature data is shown in \Fref{fig:TICS of 1,1,1,2-Tetrafluoroethane}.

\begin{table*}
\caption{\label{tab:1112tfe} The Appearance Energy (AE), Branching ratios (BR), cross section's maximum value ($\sigma$), the values of pre-exponent (A), scaling term $(n)$ and the activation energy $(\rm E_{act})$ of 1,1,1,2 Tetrafluoroethane ($\rm CF_3CH_2F$)}

\begin{tabular*}{\textwidth}{@{}l*{15}{@{\extracolsep{0pt plus
12pt}}l}}
\br
$m/z$&Cation& AE (eV)\cite{pereira2021electron}& BR&$ \sigma$ (10$^{-16}$cm$^2$)&A(10$^{-10}$cm$^3$s$^{-1}$ eV$^{-n}$)&$n$ & E$_{act}$ (eV)\\
\mr
101&CF$_3$CHF&14.56 $\pm$ 0.17&0.01 796&0.1217&0.4174& 1.6013&15.013\\
83&CF$_{3}$CH$_{2}$&15.51 $\pm$ 0.11&0.10 778&0.7227&3.120&1.513&15.7732\\
69&CF$_{3}$&13.39 $\pm$ 0.10&0.59 880&4.0765&11.51&1.6793&14.2757\\
63&CHCF$_{2}$&16.99 $\pm$ 0.25&0.02 395&0.1607&1.005&1.3664&17.0944\\
51&CHF$_{2}$&14.03 $\pm$ 0.64&0.02 395&0.1626&0.5018&1.6434&14.6414\\
33&CH$_{2}$F&13.36 $\pm$ 0.17&0.15 568&1.0600&2.984&1.6802&14.262\\
102&\textbf{CF$_3$CH$_2$F}&\textbf{13.10 $\pm$ 0.17} &$-$&7.0535&34.29&1.4986&13.3273 \\
\br
\end{tabular*}
\end{table*}
\begin{figure*}
    \centering
    \begin{tikzpicture}
    \node at (0,0) {\includegraphics[width=\textwidth,trim={0 40cm 0 0},clip]{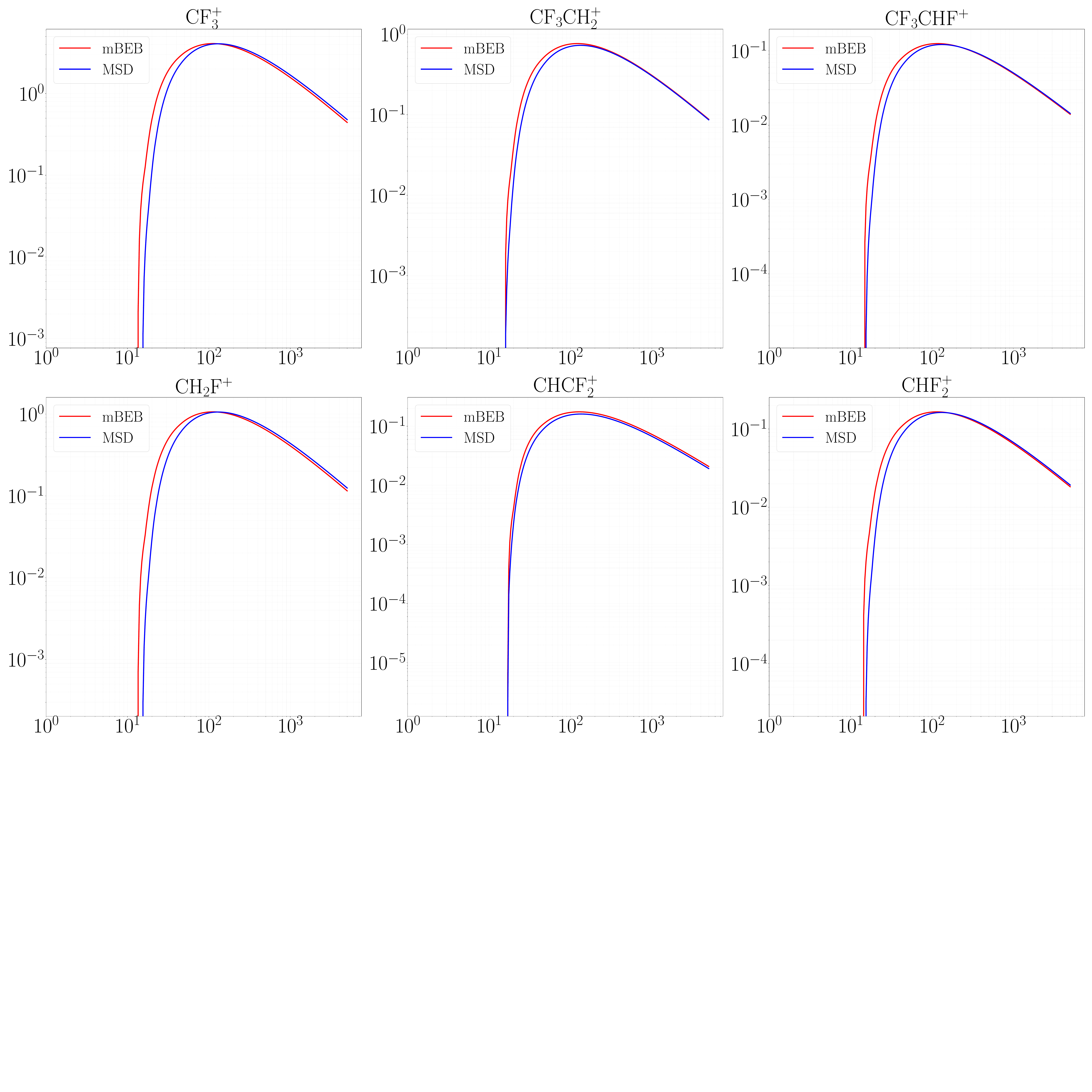}};

    \node[below, yshift=-0.5ex] at (current bounding box.south) {Energy (eV)};

    \node[left, rotate=90, xshift=9.5ex] at (current bounding box.west) {PICS $(10^{-16}cm^{2})$};
  \end{tikzpicture}
    \caption{PICS of 1,1,1,2-Tetrafluoroethane: cations detected in the EIMS, the blue solid line represents the calculated cross sections using the mBEB model, and the solid red line represents the calculated cross sections using the MSD method}
    \label{fig:PICS of 1,1,1,2-Tetrafluoroethane}
\end{figure*}

\subsection{1,1,1-Trifluoroethane}
\begin{figure*}    
    \begin{tikzpicture}
    \node at (0,0) {\includegraphics[width=\textwidth]{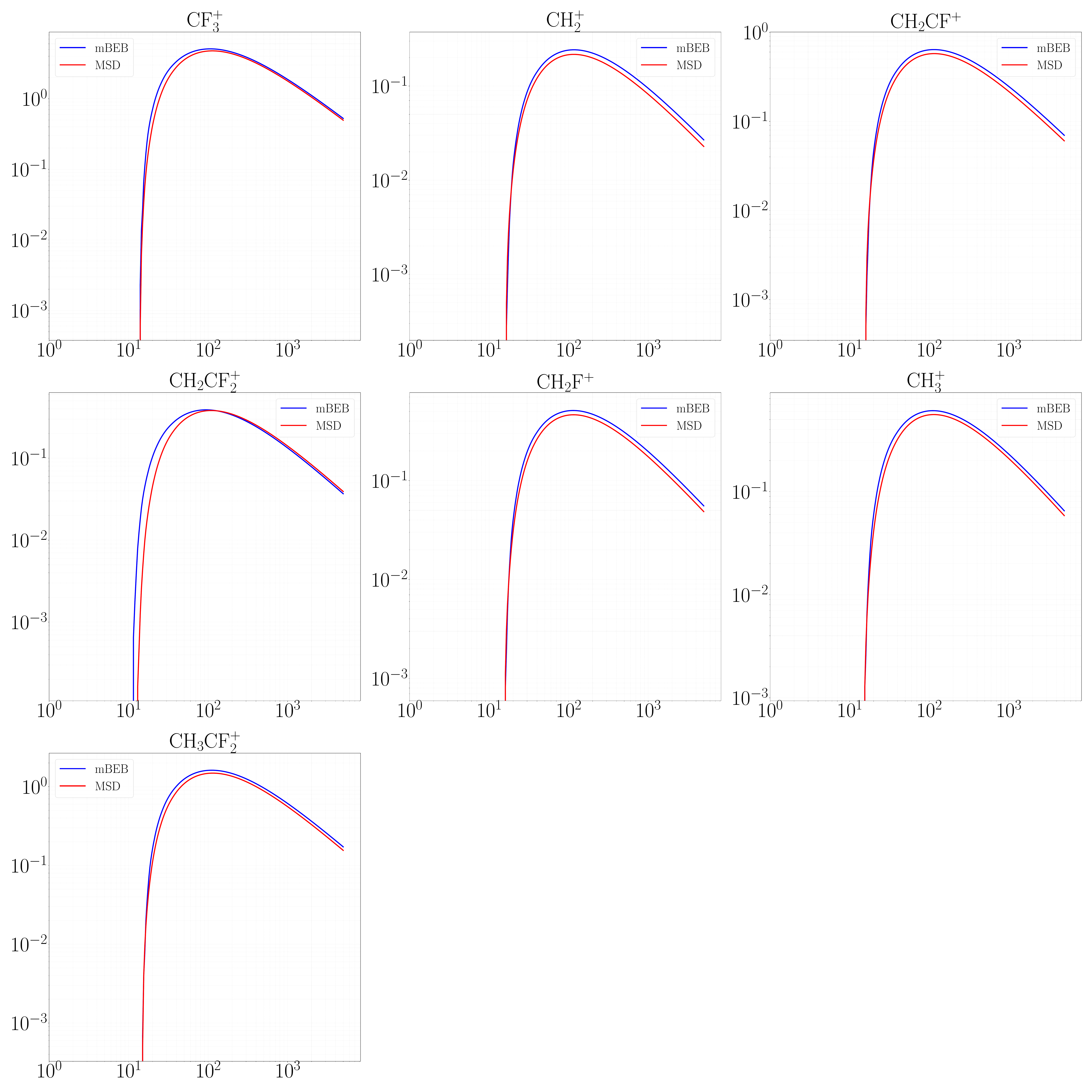}};
    
    \node[below, yshift=-0.5ex] at (current bounding box.south) {Energy (eV)};
    \node[left, rotate=90, xshift=9.5ex] at (current bounding box.west) {PICS $(10^{-16}cm^{2})$};
  \end{tikzpicture}
    \caption{\label{fig:111TFE-pics} PICS of 1,1,1-Trifluoroethane: the cations detected in the EIMS, the blue solid line represents the calculated cross sections using the mBEB model, and the solid red line represents the calculated cross sections using the MSD method.}
\end{figure*}

\begin{figure*}

\begin{subfigure}[t]{0.5\textwidth}
    \centering
    \includegraphics[width=\textwidth]{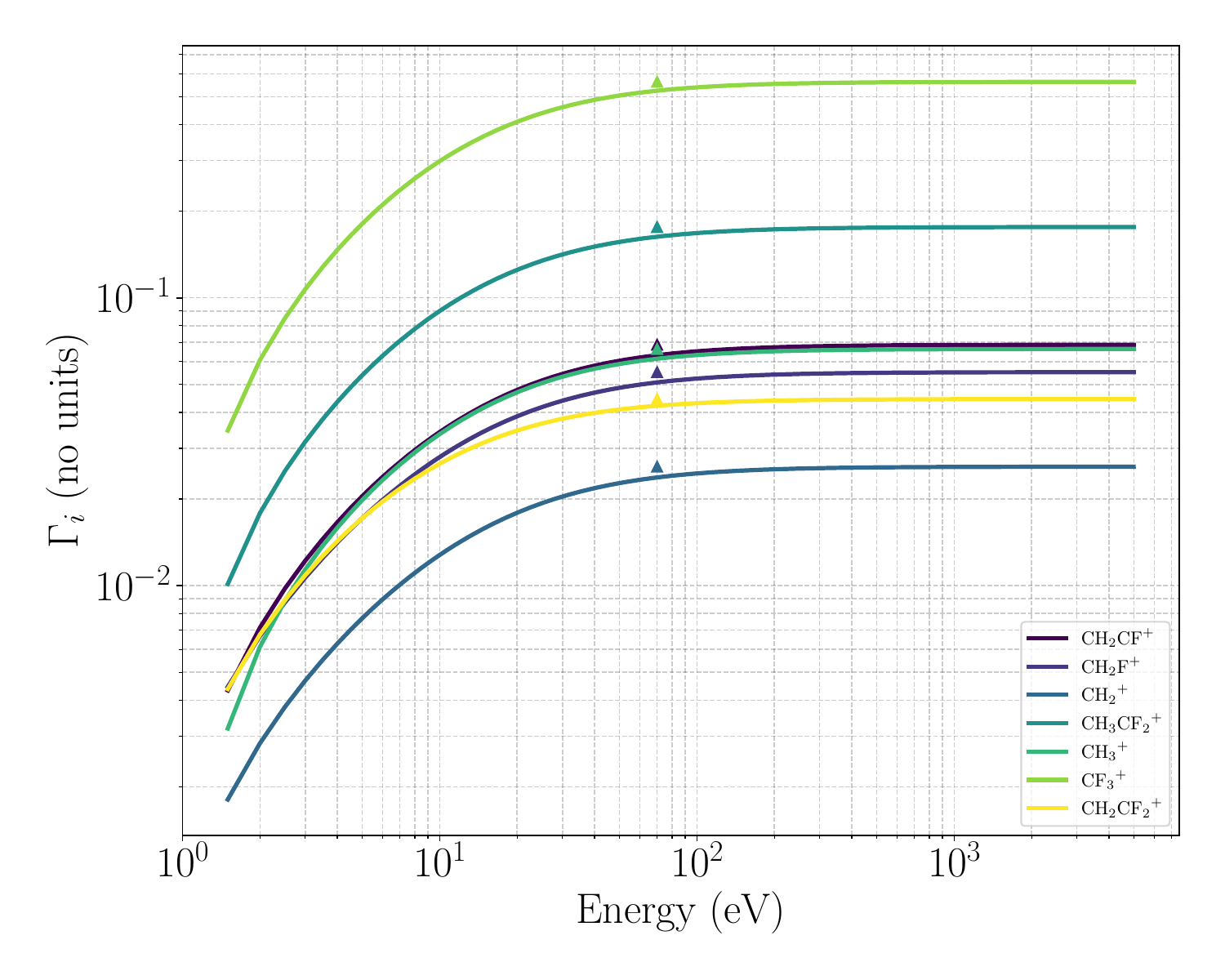}
    \caption{}
    \label{fig:111tfe-br}
\end{subfigure}
\hfill
\begin{subfigure}[t]{0.5\textwidth}
    \centering
    \includegraphics[width=\textwidth]{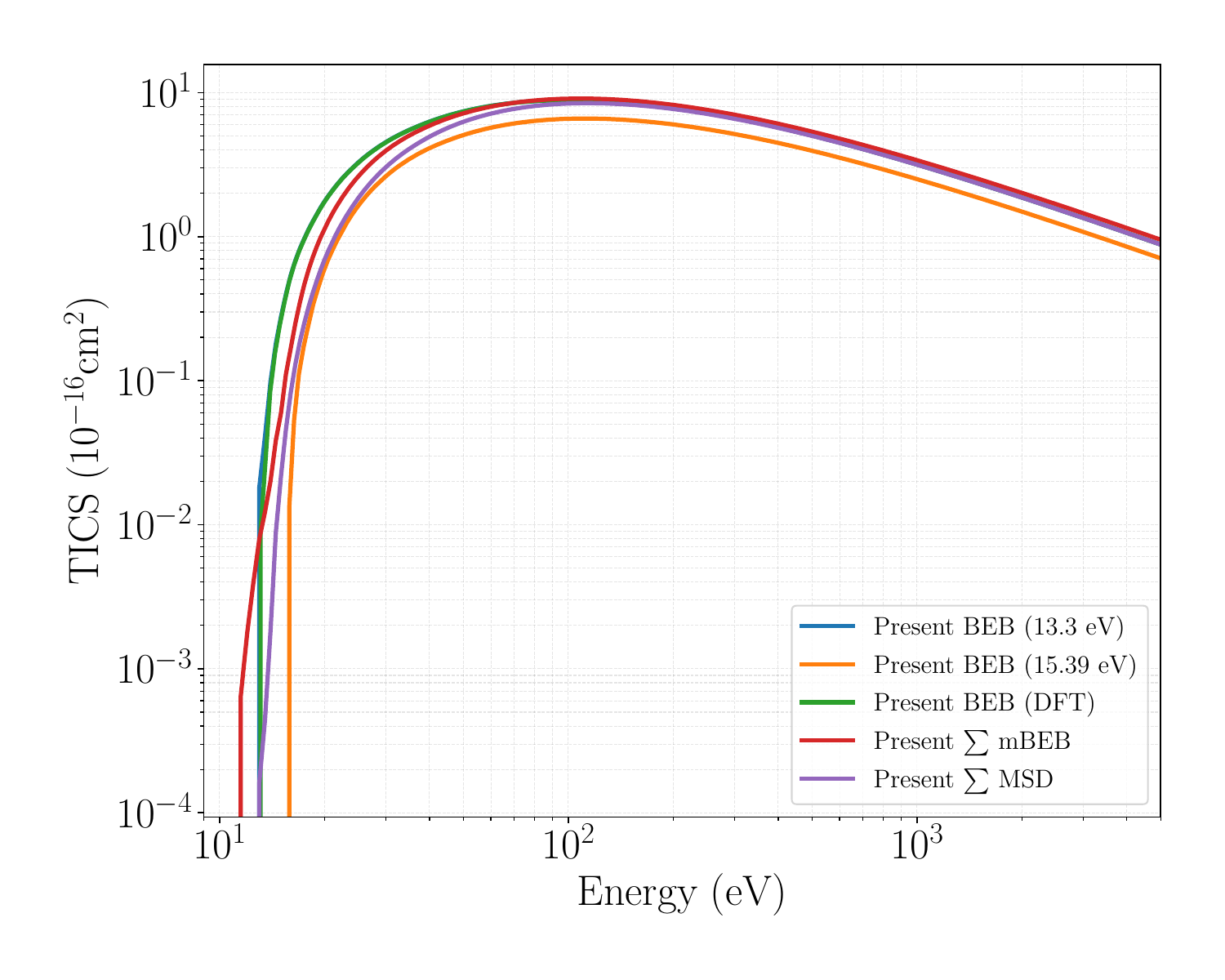}
    \caption{}
    \label{fig:111TFE-tics}
\end{subfigure}
 \hfill
\begin{subfigure}[c]{0.5\textwidth}
    \centering
    \includegraphics[width=\textwidth]{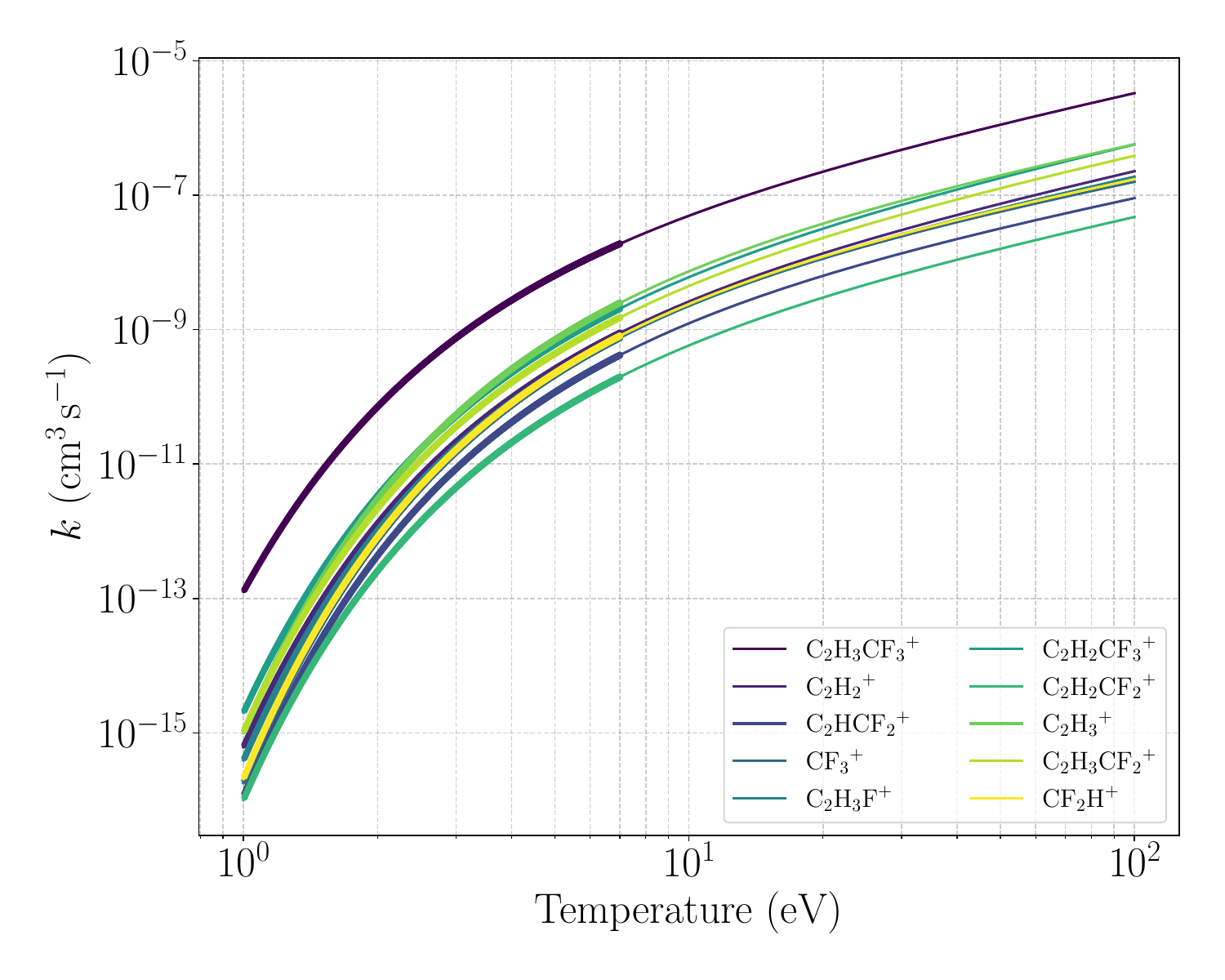}
    \caption{}
    \label{fig:rates-111tfe-rates}
    \end{subfigure}
    \caption{a) Branching ratios of 1,1,1-Trifluoroethane (CH$_3$CF$_3$): The solid lines represent the branching ratios calculated for the MSD method, whereas the solid triangles denote the branching ratios calculated from the EIMS. 
    b) Comparison of TICS of 1,1,1-Trifluoroethane calculated using different methods, the solid blue line represents our calculated BEB TICS where the experimental first ionization energy was used, the solid orange line shows the present BEB TICS calculated using HOMO energy obtained by HF approximation, the green line represents the TICS calculated using the DFT parameters, the red line represents the sum of mBEB PICS, the purple line represents the sum of PICS calculated using the MSD method. 
    c) 1,1,1-Trifluoroethane rate coefficients for ionization and dissociation processes, here the bold lines are the rates calculated using the MVD, and the thin solid lines are the extrapolated values calculated using the Arrhenius function from 1 eV to 100 eV.}
\end{figure*}

In the case of 1,1,1-Trifluoroethane, the mass spectrum and appearance energies were obtained from the work of Steele and Stone \cite{steele1962electron}. The AEs were measured using the modified Consolidated Electrodynamics Corporation (CEC) 21-103 C mass spectrometer. Later Simme and Tschuikow-Roux \cite{simmie1971mass} also measured the mass spectrum using the Varian Atlas CH-5 spectrometer and the appearance potential (AP) of the 1,1,1 Trifluoroethane molecule was calculated from the semilog plots of the iron current vs the electron voltage. This work has revisited the experiment and has measured the mass spectrum of three more important cations that Steele and Stone did not detect; this includes fragments such as $m/z:$ Cation $-$ 84: CH$_3$CF$_3$, 63: CHCF$_2$, 51: CHF$_2$ and 31: CF. On the contrary, all these cations have relative abundances (RI) $<5\%$. A cation of $m/z$ of 14: CH$_2$ went undetected in their measurement. This cation was then detected by Steele and Stone which had RI of $<5\%$. The shortcomings in the study of Simme and Tschuikow-Roux is that the APs were not measured for all the cations. Hence we have used the data provided by Steele and Stone \cite{steele1962electron} for RI and AEs of the cations. In \Tref{tab:tfe}, we have shown the calculated BRs along with the PICS and the fitting parameters of the rate constants. In \Fref{fig:111TFE-pics}, we have presented the PICS for the seven cationic fragments detected by Steele and Stone \cite{steele1962electron}. As the AEs were not provided for CH$_3$CF$_3^+$, CHCF$_2^+$, CHF$_2^+$, and CF$^+$, we could not calculate the PICS for these fragments that may be of interest as they induce further ion-ion, ion-atom, and ion-molecule processes in dense environments. We would also prescribe revisiting the study with the current instrumentation as they are quite sensitive and precise and would also help us determine characteristics of the smaller fragments ($<20$ amu) including the C$_2^+$, C$^+$ and H$_2^+$ and the larger fragments such as the CF$_x^+(x = 2,3)$. In \Fref{fig:111tfe-br}, we have shown the branching ratios as a function of incident energy (E), the solid lines represent the continuous branching ratios used in the MSD method and the upright triangles are the BRs from the EIMS data, measured at a particular incident energy of 70 eV. In \Fref{fig:111TFE-tics}, and \Fref{fig:rates-111tfe-rates} we have shown the TICS calculated using orbital parameters calculated using the $\mathcal{HF}$ and the DFT level of theory. As our binding energy calculations using the $\mathcal{HF}$ method did not yield an ionization energy\footnote{According to Koopman's theorem IE = - HOMO} closer to the experimental value or the value reported in the literature, we replaced our HOMO energy (15.39 eV) to 13.3 eV, which was reported in the NIST web book. The DFT calculation yielded a HOMO energy value of 12.58 eV which is closer to the value in the NIST Web book. This difference in the HOMO energy affects the magnitude and the starting point (ionization threshold) of the TICS profiles, which can be seen in \Fref{fig:111TFE-tics}. A similar trend can be noticed in the TICS calculated from the sum of PICS obtained from the mBEB and MSD methods, since for a few cations, their appearance energies are lesser than the ionization energies.  In \Fref{fig:rates-111tfe-rates}, we have shown the rates calculated using Maxwell's electron velocity distribution (MVD) function from 1 eV till 7 eV and fitted using the Arrhenius function from 1 eV till 100 eV, the fitting constants such as A, $n$ and E$_{act}$ have been presented in \Tref{tab:tfe}. 

\begin{table*}
\caption{\label{tab:tfe} The Appearance Energy (AE), Branching ratios (BR), cross section's maximum value ($\sigma$), the values of pre-exponent (A), scaling term $(n)$ and the activation energy $(\rm E_{act})$ of 1,1,1 Trifluoroethane (CH$_3$CF$_3$) }

\begin{tabular*}{\textwidth}{@{}l*{15}{@{\extracolsep{0pt plus
12pt}}l}}
\br
$m/z$&Cation& AE (eV)& BR&$ \sigma$ (10$^{-16}$cm$^2$)&A(10$^{-10}$cm$^3$s$^{-1}$ eV$^{-n}$)&$n$ & E$_{act}$ (eV)\\
\mr
69&CF$_3$&13.9 $\pm$ 0.03 &0.56 306&3.6606&18.34&1.4678&14.7157
\\
65&CH$_3$CF$_2$&14.9 $\pm$ 0.2 &0.17 623&1.1407&6.806&1.3972
&15.3955\\
64&CH$_2$CF$_2$&11.2 $\pm$ 0.1 &0.04 448&0.2924&1.330&1.513
&13.7573\\
45&CH$_2$CF&15.8 $\pm$ 0.2&0.06 869&0.4428&3.060&1.3399
&16.0405\\
33&CH$_2$F&15.6 $\pm$ 0.2&0.05 518&0.3560&2.391&1.3508
&15.8997\\
15&CH$_3$&15.0 $\pm$ 0.1&0.06 644&0.4298&2.628&1.3875
&15.4783\\
14&CH$_2$&16.2 $\pm$ 0.3&0.02 590&0.1666&1.233&1.3138
&16.3484\\
84&\textbf{CH$_3$CF$_3$}&\textbf{13.3 $\pm$ 0.1}&$-$&6.7843&52.16&1.3415
&13.3577\\
\br

\end{tabular*}
\end{table*}

\subsection{1,1,1-Trifluoropropane \& 3,3,3-Trifluoropropene}
\begin{figure*} 
\begin{subfigure}[t]{0.5\textwidth}
    \centering
    \includegraphics[width=\textwidth]{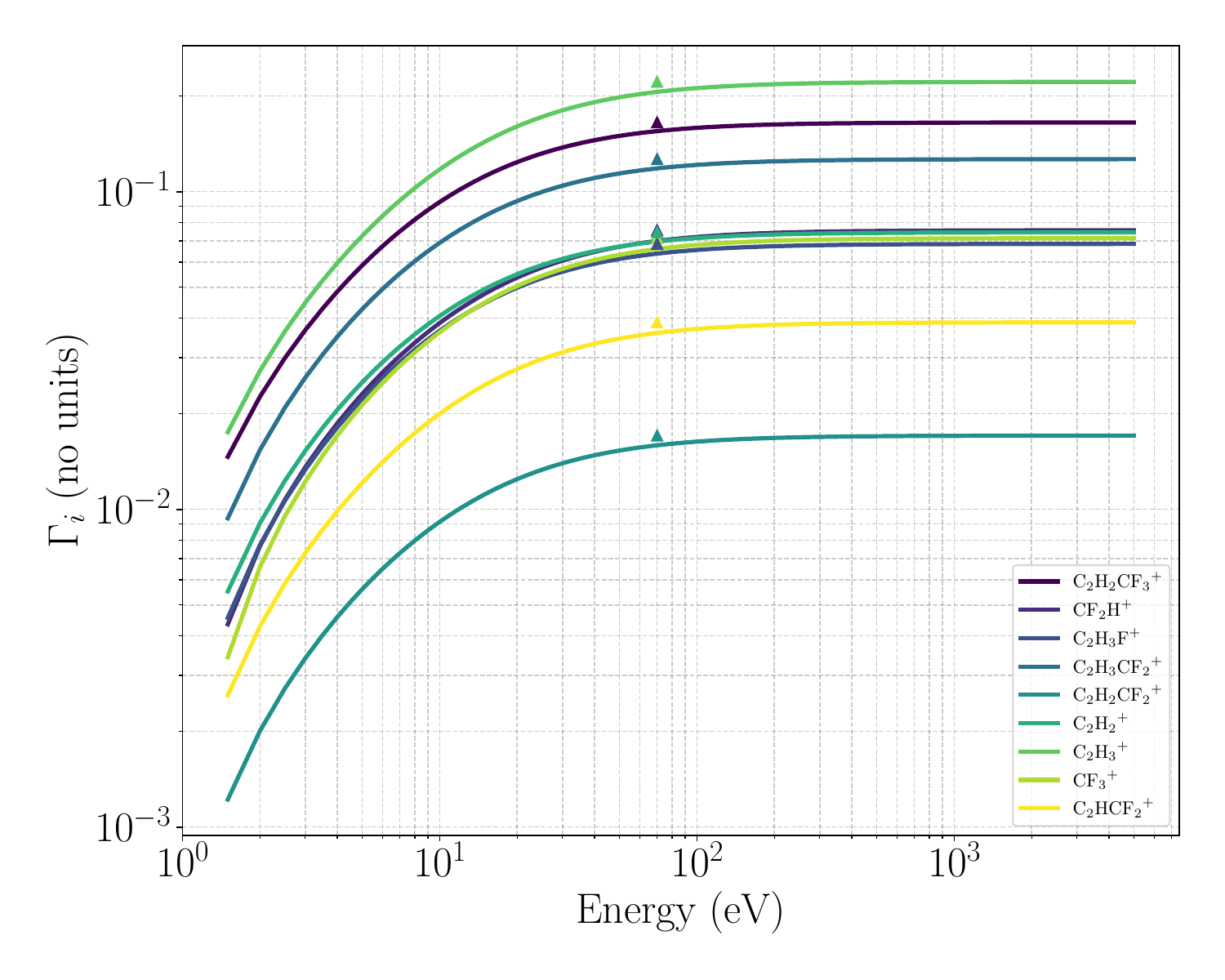}
    \caption{}
    \label{fig:111tfp-br}
\end{subfigure}
\begin{subfigure}[t]{0.5\textwidth}
    \centering
    \includegraphics[width=\textwidth]{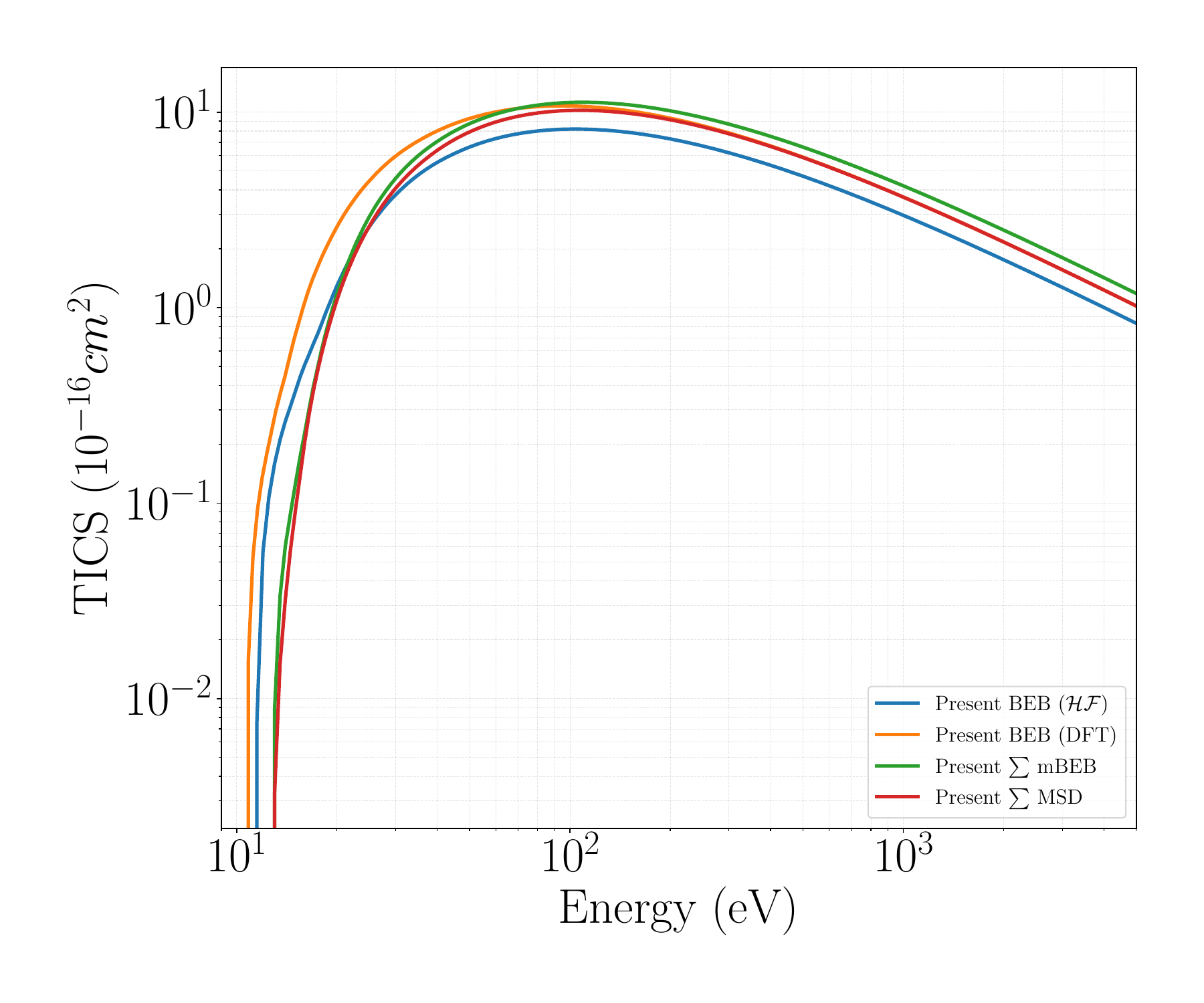}
    \caption{}
    \label{fig:TICS-1,1,1-Trifluoropropane}
\end{subfigure}

\begin{subfigure}[b]{0.5\textwidth}
    \centering
    \includegraphics[width=\textwidth]{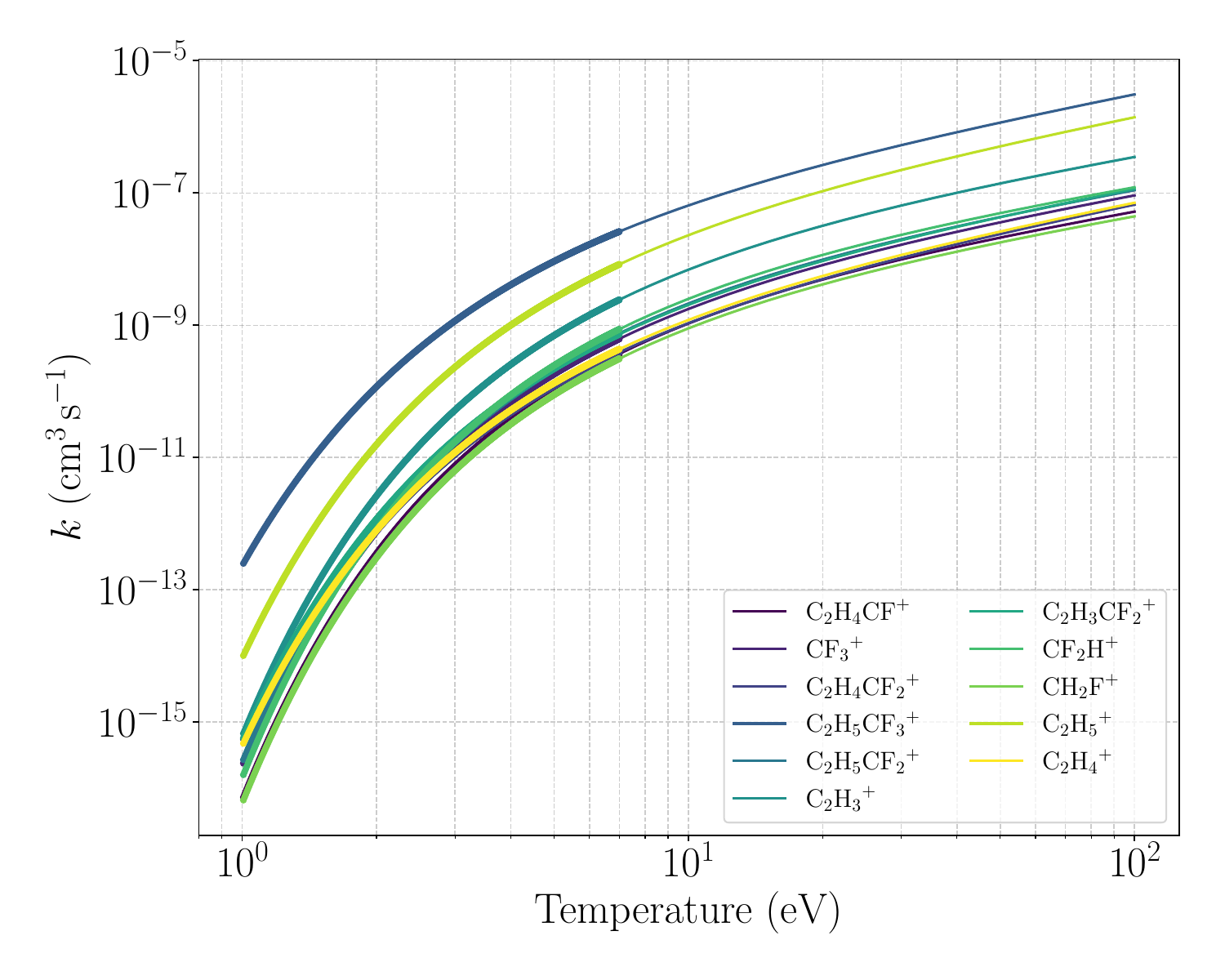}
    \caption{}
    \label{fig:111tfp-rates}
\end{subfigure}
\caption{a) Branching ratios of 1,1,1-Trifluoropropane (C$_2$H$_5$CF$_3$): The solid lines represent the branching ratios calculated for the MSD method, whereas the solid triangles denote the branching ratios calculated from the EIMS. b) 1,1,1-Trifluoropropane: TICS calculated using different methods, the solid blue line represents our calculated BEB TICS calculated using HOMO energy obtained by HF approximation, and the solid orange line shows the present BEB TICS calculated using orbital energies obtained by DFT method, the solid green line represents the sum of all PICS calculated from the mBEB method, the solid red line represents the PICS calculated using the MSD method. c) 1,1,1-Trifluoropropane rate coefficients for ionization and dissociation processes, here the bold lines are the rates calculated using the MVD, and the thin solid lines are the extrapolated values calculated using the Arrhenius function from 1 eV to 100 eV.}
\end{figure*}

1,1,1-Trifluoropropane molecule's mass spectrum and AEs were also obtained from the work of Steele and Stone \cite{steele1962electron}. There are no previous reports on ionization potential along with ionization cross  sections studies  on the molecule in the literature to the best of our knowledge.  From \Tref{tab1-IE}, it can be seen that our HOMO energy calculated using $\mathcal{HF}$ method is 11.42 eV and that calculated using DFT method is 10.69 eV. The calculated BRs, PICS, and fitting parameters are shown in \Tref{tab:111tfp}. In \Fref{fig:PICS-1,1,1-Trifluoropropane}, we have shown the PICS for the cations detected in the mass spectrum. The C$_2$H$_5^+$ had the most intense peak having an AE of 12.82 eV. The BRs have also been shown on \Fref{fig:111tfp-br}. The TICS calculated using the parameters obtained from the $\mathcal{HF}$ and the DFT method have been plotted in \Fref{fig:TICS-1,1,1-Trifluoropropane}, and the \Fref{fig:111tfp-rates}, the fitting parameters have been shown in \Tref{tab:111tfp}.

\begin{table*}
\caption{\label{tab:111tfp} The Appearance Energy (AE), Branching ratios (BR), cross section's maximum value ($\sigma$), the values of pre-exponent (A), scaling term $(n)$ and the activation energy $(\rm E_{act})$ of 1,1,1 Trifluoropropane ($\rm{C_2H_5CF_3}$)}

\begin{tabular*}{\textwidth}{@{}l*{15}{@{\extracolsep{0pt plus
12pt}}l}}
\br
$m/z$&Cation& AE (eV)& BR&$ \sigma$ (10$^{-16}$cm$^2$)&A(10$^{-10}$cm$^3$s$^{-1}$ eV$^{-n}$)&$n$ & E$_{act}$ (eV) \\

\mr
79&C$_2$H$_5$CF$_2$&14.9 $\pm$ 0.2&0.05 369&0.4180&6.952&1.1321
&14.8573\\
78&C$_2$H$_4$CF$_2$&12.53 $\pm$ 0.04&0.02 330&0.1834&2.007&1.2886
&12.8908\\
77&C$_2$H$_3$CF$_2$&13.6 $\pm$ 0.1&0.04 711&0.3690&4.892&1.2154&13.7399\\
69&CF$_3$&14.8 $\pm$ 0.1 &0.04 407&0.3432&5.582&1.1404&14.7586
\\
59&C$_2$H$_4$CF&15.8 $\pm$ 0.1&0.02 836&0.2199&4.249&1.0788&15.6748

\\
51&CF$_2$H&15.9 $\pm$  0.1&0.06 737&0.5220&10.32&1.0705&15.7796\\
33&CH$_2$F&15.7 $\pm$ 0.3&0.02 380&0.1846&3.511&1.0844&15.5825
\\
29&C$_2$H$_5$&12.82 $\pm$ 0.02&0.50 658&3.9826&45.60&1.2712&13.1025
\\
28&C$_2$H$_4$&13.0 $\pm$ 0.2&0.02 684&0.2109&2.528&1.2534&13.267\\
27&C$_2$H$_3$&15.3 $\pm$ 0.1&0.17 882&1.3895&24.64&1.1093&15.2106\\
98&\textbf{C$_2$H$_5$CF$_3$}&$\textbf{11.42}$&$-$&8.1920&103.3&1.2614&10.6984\\
\br

\end{tabular*}
\end{table*}

\begin{figure*}
    \centering
    \begin{tikzpicture}
    \node at (0,0) {\includegraphics[width=\textwidth]{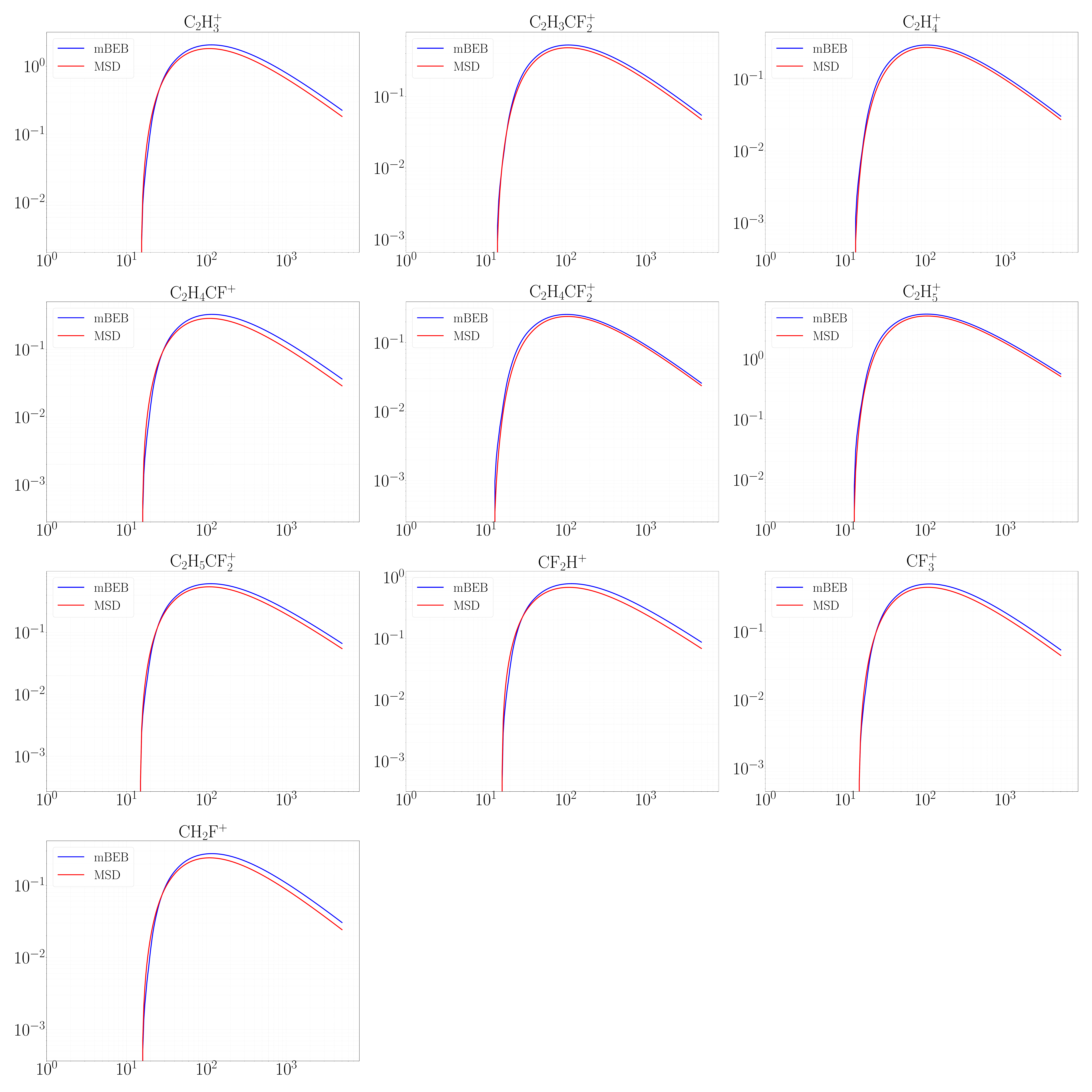}};

    \node[below, yshift=-0.5ex] at (current bounding box.south) {Energy (eV)};

    \node[left, rotate=90, xshift=9.5ex] at (current bounding box.west) {PICS $(10^{-16}cm^{2})$};
  \end{tikzpicture}
    \caption{ PICS of 1,1,1-Trifluoropropane:  cations detected in the EIMS, the blue solid line represents the calculated cross sections using the mBEB model, and the solid red line represents the calculated cross sections using the MSD method.}
    \label{fig:PICS-1,1,1-Trifluoropropane}
\end{figure*}

\begin{figure*}
\begin{subfigure}[t]{0.5\textwidth}
    \centering
    \includegraphics[width=\textwidth]{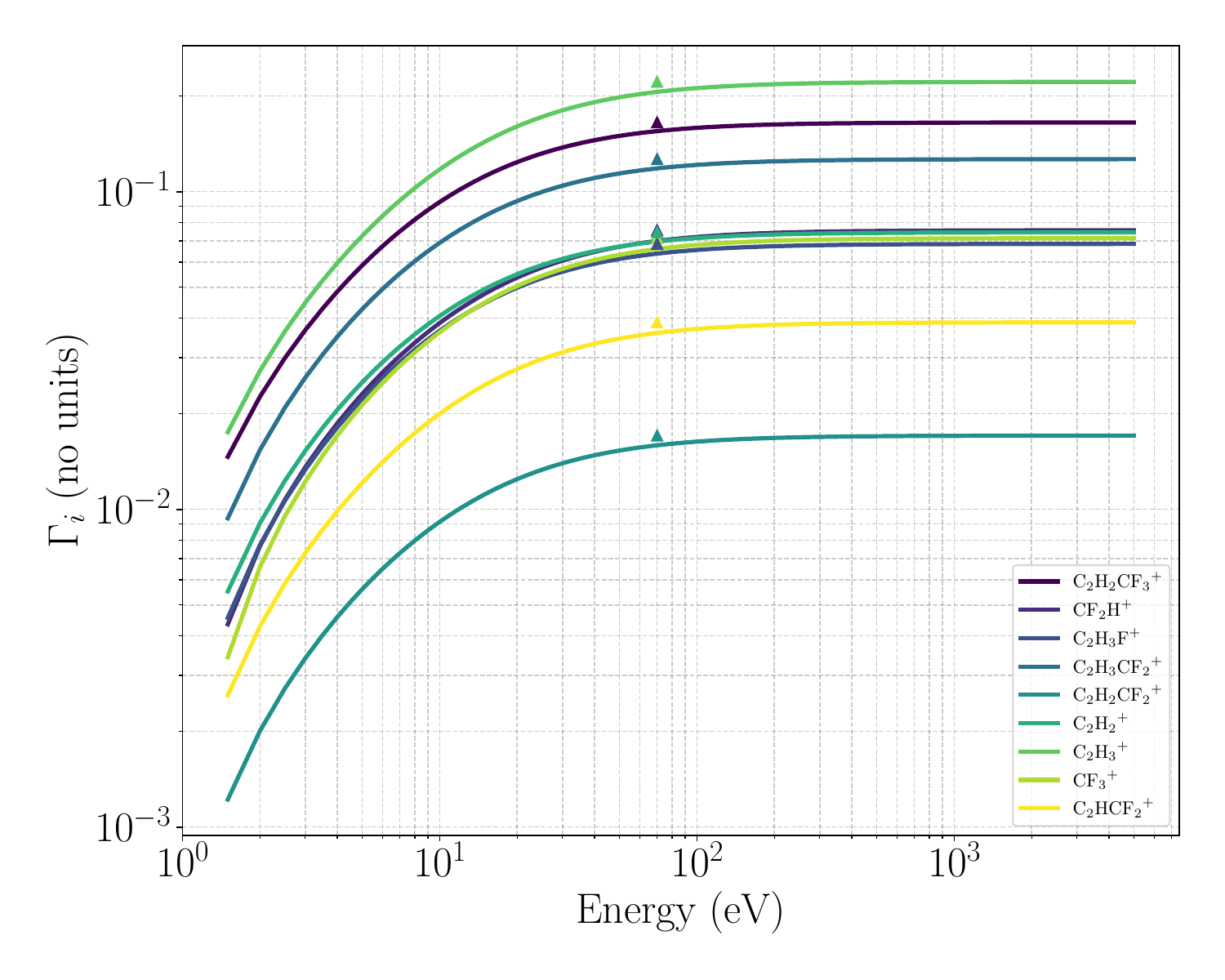}
    \caption{}
    \label{fig:333tfp-br}
\end{subfigure}
\begin{subfigure}[t]{0.5\textwidth}
    \centering
    \includegraphics[width=\textwidth,height=0.83\textwidth]{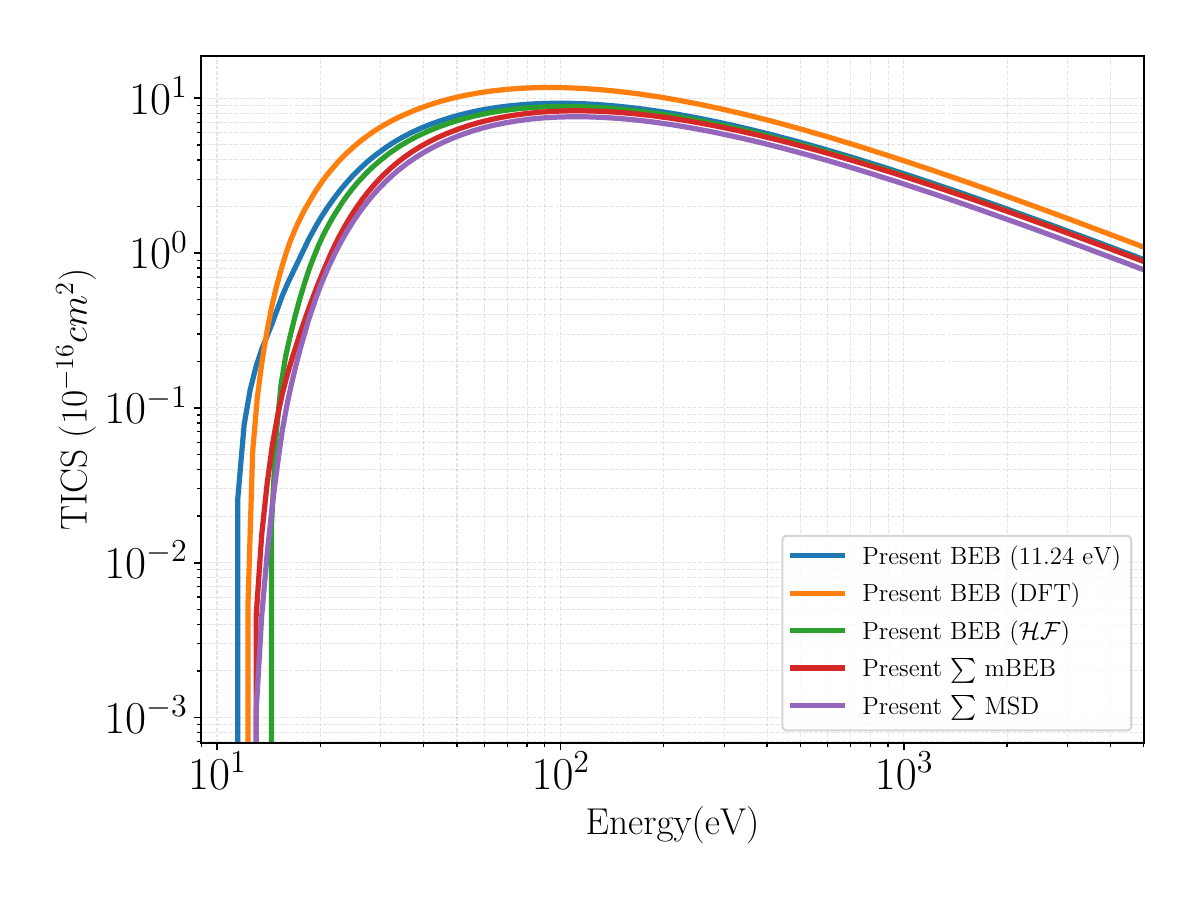}
    \caption{}
    \label{fig:TICS-3,3,3-Trifluoropropene}
\end{subfigure}
\begin{subfigure}[t]{0.5\textwidth}
    \centering
    \includegraphics[width=\textwidth]{Figures/Rates/C2H3CF3_Rates.pdf}
    \caption{}
    \label{fig:rates-333tfp}
\end{subfigure}
\caption{a) Branching ratios of 3,3,3-Trifluoropropene C$_3$H$_3$CF$_3$: The solid lines represent the branching ratios calculated for the MSD method, whereas the solid triangles denote the branching ratios calculated from the EIMS. b) 3,3,3-Trifluoropropene: TICS calculated using different methods, the solid blue line represents our calculated BEB TICS where the experimental first ionization energy was used, and the solid orange line shows the present BEB TICS calculated using values obtained from the DFT method, the solid green line shows the present BEB TICS calculated using HOMO energy obtained by HF approximation, the red lines represents the sum of PICS calculated from the mBEB method, and the purple lines show the sum of PICS calculated from the MSD method. c) 3,3,3-Trifluoropropene rate coefficients for ionization and dissociation processes, here the bold lines are the rates calculated using the MVD, and the thin solid lines are the extrapolated values calculated using the Arrhenius function from 1 eV to 100 eV. }
\end{figure*}

Similarly, 3,3,3-Trifluoropropene was also studied by Stone and Steele, and mass spectrum and AEs were provided \cite{steele1962electron}. The reported IE is 11.24 $\pm$ 0.04 eV. Our calculated HOMO values by the DFT and $\mathcal{HF}$ methods are 12.16 eV and 14.22 eV. The \Fref{fig:PICS-3,3,3-Trifluoropropene}, contains the PICS calculated for all the cationic fragments presented in \Tref{tab:333tfp}. Here the C$_2$H$_3^+$ is the most abundant fragment having an AE of 14.20 eV. The parent fragment C$_2$H$_3$CF$_3^+$ was also detected in the mass spectrum having an RI $>60\%$. The BR has been shown in \Fref{fig:333tfp-br}, where the solid lines represent the continuous BRs and the upright triangle represents the experimental BR at 70 eV. 
The TICS calculated using the parameters from the various methods have been shown in the \Fref{fig:TICS-3,3,3-Trifluoropropene}, the HOMO energy obtained from the $\mathcal{HF}$ method is quite high compared to the experimental values hence there is a shift near the threshold of the cross sections, whereas when we replace the HOMO energy value with parameters obtained by DFT method, the cross section profile begins at a lower threshold value and a higher magnitude when compared with the TICS calculated using the $\mathcal{HF}$ method. The TICS obtained by summing all the PICS of the cations from the MSD and the mBEB methods are in good agreement with the BEB TICS calculated using the $\mathcal{HF}$ approximation for orbital parameters. The rate coefficients for the PICS and TICS are shown in \Fref{fig:rates-333tfp}, the rates were calculated using the MVD from 1 eV till 7 eV and fitted using the Arrhenius function from 1 eV till 100 eV.


\begin{table*}
\caption{\label{tab:333tfp} The Appearance Energy (AE), Branching ratios (BR), cross section's maximum value ($\sigma$), the values of pre-exponent (A), scaling term $(n)$ and the activation energy $(\rm E_{act})$ of 3,3,3 Trifluoropropane ($\rm{C_2H_3CF_3}$)}
\begin{tabular*}{\textwidth}{@{}l*{15}{@{\extracolsep{0pt plus
12pt}}l}}
\br
$m/z$&Cation& AE (eV)& BR&$ \sigma$ (10$^{-16}$cm$^2$)&A(10$^{-10}$cm$^3$s$^{-1}$ eV$^{-n}$)&$n$ & E$_{act}$ (eV) \\
\mr 
95&C$_2$H$_2$CF$_3$&12.69 $\pm$ 0.05&0.16 515&1.4696&7.322&1.4747&12.8115\\
77&C$_2$H$_3$CF$_2$&13.3 $\pm$  0.15&0.12 657&1.1230&6.487&1.4163&13.3467\\
76&C$_2$H$_2$CF$_2$&13.8 $\pm$ 0.1& 0.01 706&0.1510&98.24&1.3716&13.7944\\
75&C$_2$HCF$_2$&14.8 $\pm$ 0.2&0.03 879&0.3416&2.785&1.2877&14.7085\\
69&CF$_3$&15.0 $\pm$ 0.2&0.07 138&0.6279&5.444&1.2644&14.9268\\
51&CF$_2$H&14.9 $\pm$ 0.1&0.07 559&0.6653&5.573&1.2774&14.8112
\\
46&C$_2$H$_3$F& 13.85 $\pm$ 0.02&0.06 849&0.6061&3.994&1.3666&13.8423\\
27&C$_2$H$_3$&14.20 $\pm$ 0.05&0.22 168&1.9580&13.97&1.337&14.1573\\
26&C$_2$H$_2$&13.3 $\pm$ 0.15&0.07 448&0.6608&3.818&1.4163&13.3467
\\
96&\textbf{C$_2$H$_3$CF$_3$}&\textbf{11.24 $\pm$ 0.04}&$-$&9.2915&56.96&1.4043&10.7188\\
\br

\end{tabular*}
\end{table*}

\begin{figure*}
    \centering
    \begin{tikzpicture}
    \node at (0,0) {\includegraphics[width=\textwidth]{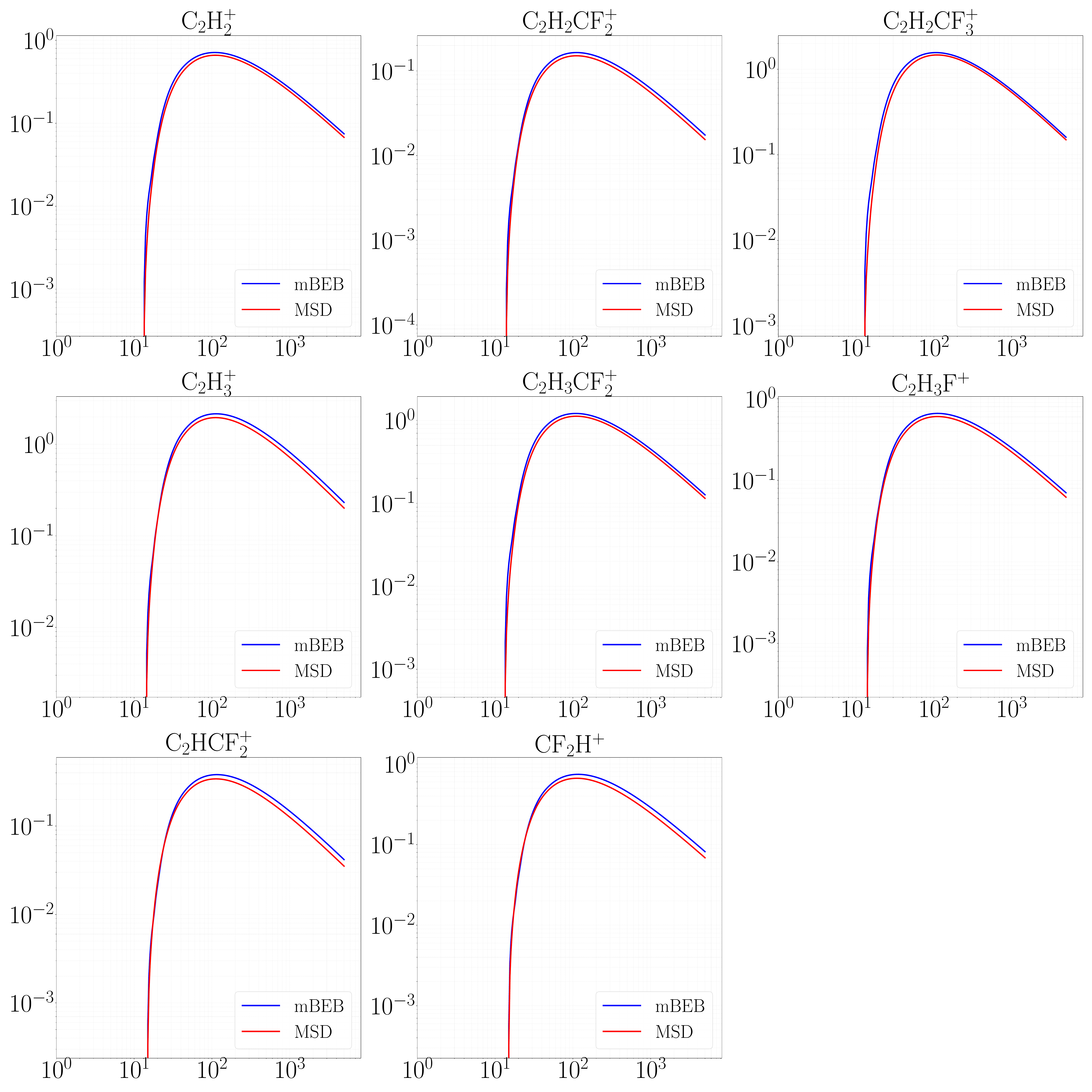}};

    \node[below, yshift=-0.5ex] at (current bounding box.south) {Energy (eV)};

    \node[left, rotate=90, xshift=9.5ex] at (current bounding box.west) {PICS $(10^{-16}cm^{2})$};
  \end{tikzpicture}
    \caption{PICS of 3,3,3-Trifluoropropene: cations detected in the EIMS, the blue solid line represents the calculated cross sections using the mBEB model, and the solid red line represents the calculated cross sections using the MSD method.}
    \label{fig:PICS-3,3,3-Trifluoropropene}
\end{figure*}

\section{Conclusion}
\begin{figure}[h]
    \begin{subfigure}[t]{0.5\textwidth}
        \centering
        \includegraphics[width=\textwidth]{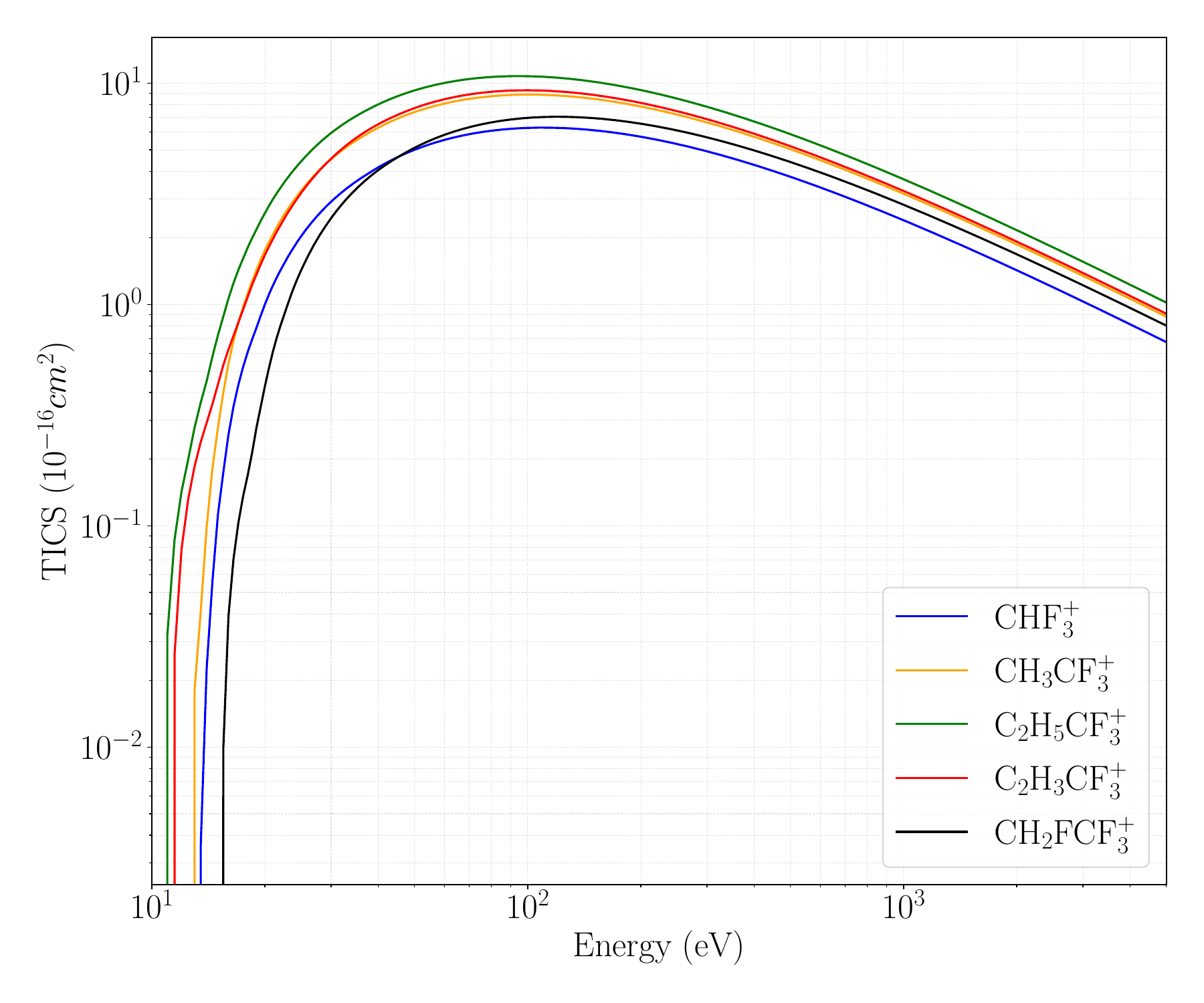}
        \caption{}       
    \end{subfigure}
    \hfill
    \begin{subfigure}[t]{0.5\textwidth}
        \centering
        \includegraphics[width=\textwidth]{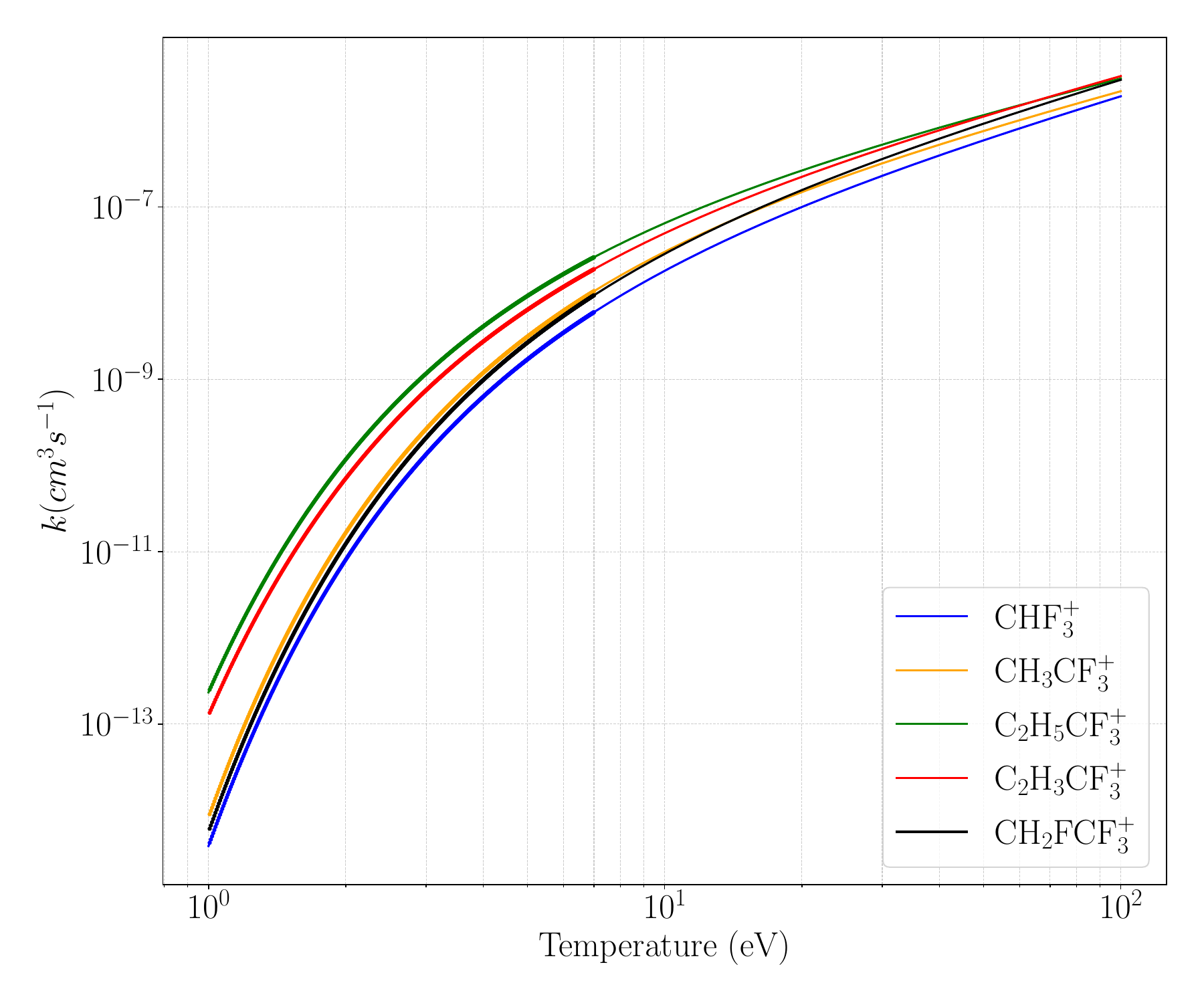}
        \caption{}
        \end{subfigure}
        \caption{The total electron impact ionization cross sections and the ionization rate constants of all the molecules discussed in the manuscript.}
        \label{last_plot}
\end{figure}

The electron impact PICS have been successfully calculated for the cations of several molecules are reported in the article. The PICS was computed using the mBEB and the MSD models. These models require the TICS, which is calculated using the BEB model. All the molecules that are studied in this article were optimised using the DFT functional ($\omega B97XD$) and the aug-cc-PVTZ basis set. The orbital binding and kinetic energies were calculated using the $\mathcal{HF}$ and DFT approximations using the same aug-cc-PVTZ basis set. Since the HOMO energies of all the molecular targets were high, we replaced them with the experimental IPs which are compared in \Tref{tab1-IE} with our HOMO energy. The computed cross sections have been used to calculate the thermal rate constants for the ionization and dissociative ionization processes. Our results for molecules like trifluoromethane and 1,1,1,2- tetrafluoroethane showed good agreement with the literature data and are quite well studied. As seen in \Fref{fig:CHF3tics}, our TICS compares very well with the experimental data of Torres et al \cite{torres2001evaluation}, Kawaguchi et \cite{kawaguchi2014electron}, the recommended cross sections of Christophorou et al \cite{christophorou2004electron}, and the theoretical calculation of Kim et al \cite{torres2001evaluation}. However, the molecules such as 1,1,1-Trifluoroethane, 1,1,1-Trifluoropropane and 3,3,3-Trifluropropene do not have any ionization studies in the literature and hence this study provides a comprehensive data for these targets for the first time. Also we have found that the agreement between the activation energy and appearance energy is linear as can be observed from \cref{tab:fit-para,tab:fluoroform,tab:tfe,tab:111tfp,tab:333tfp,tab:1112tfe}. However, no direct functional dependence or correlation is available. Our opinion is that revisiting these molecules with the current apparatus will provide a more comprehensive investigation of these targets.   Finally, a comparison of all the TICS and their total ionization rate constants have been shown in \Fref{last_plot} where normally with the size of the target the TICS increases. 

\newpage
\bibliographystyle{unsrt}
\bibliography{References}
\end{document}